\documentclass[11pt,a4paper,nofootinbib,showpacs,preprint,utf8]{article}
\usepackage{jheppub}
\pdfoutput=1
\usepackage{amsmath,amssymb}
\usepackage{amsfonts}
\usepackage{epsfig}
\usepackage{graphicx}
\usepackage{amssymb}
\usepackage{bbm}
\usepackage{pifont}
\usepackage[usenames,dvipsnames]{xcolor}
\definecolor{warmblack}{rgb}{0.0, 0.26, 0.26}
\definecolor{mediumtealblue}{rgb}{0.0, 0.33, 0.71}
\usepackage[compat=1.1.0]{tikz-feynman}
\usepackage{tikzsymbols}
\usepackage{comment}
\usepackage[normalem]{ulem}
\usepackage{orcidlink}

\title{Gravitational wave signatures of dark sector portal leptogenesis}
\author[a,b]{{Debasish Borah}}
\author[c]{{, Devabrat Mahanta}}
\author[a]{{, Indrajit Saha}}

\affiliation[a]{Department of Physics, Indian Institute of Technology Guwahati, Assam 781039, India}
\affiliation[b]{Pittsburgh Particle Physics, Astrophysics, and Cosmology Center, Department of Physics and Astronomy, University of Pittsburgh, Pittsburgh, PA 15260, USA}
\affiliation[c]{Department of Physics, Pragjyotish College, Guwahati, Assam 781009, India}

\emailAdd{dborah@iitg.ac.in}
\emailAdd{devabrat@pragjyotishcollege.ac.in}
\emailAdd{s.indrajit@iitg.ac.in}

\abstract{We study the possibility of probing leptogenesis via stochastic gravitational waves (GW) arising from a dark sector assisted first-order electroweak phase transition. The same dark sector, with non-trivial transformation under an unbroken $Z_2$ symmetry is also responsible for providing the only source of CP asymmetry via one-loop interference with the tree level decay of a heavy right-handed neutrino into lepton and Higgs doublets. The new Yukawa and scalar portal couplings enhance the CP asymmetry allowing TeV scale leptogenesis without any resonant enhancement. Light neutrino masses arise from a combination of type-I and one-loop contributions with vanishing lightest neutrino mass. While the new degrees of freedom in sub-TeV range keep the detection prospects at terrestrial experiments promising, the new scalars enhance the strength of the electroweak phase transition keeping the GW signals within reach of near future experiments like LISA.}

\makeatletter
\gdef\@fpheader{}
\makeatother

\begin{document}
\maketitle

\section{Introduction}
Several measurements in astrophysics and cosmology suggest the presence of dark matter (DM) and baryon asymmetry in the Universe (BAU) \cite{Zyla:2020zbs, Aghanim:2018eyx}. While DM corresponds to approximately $27\%$ of the present Universe $(\Omega_{\rm DM}h^2 = 0.12)$, the visible or baryonic sector contributes only around $5\%$ to the present energy density. The observed BAU is quantified in terms of the baryon to photon ratio given by \cite{Aghanim:2018eyx} 
\begin{equation}
\eta_B = \frac{n_{B}-n_{\bar{B}}}{n_{\gamma}} = 6.1 \times 10^{-10}\,,
\label{etaBobs}
\end{equation}
consistent with both cosmic microwave background (CMB) measurements and successful predictions of the big bang nucleosynthesis (BBN). The standard model (SM) of particle physics, in spite of its phenomenological succeess, fails to provide any explanations for these two observed phenomena, leading to longstanding puzzles in particle physics and cosmology. While the SM does not have a particle DM candidate, it also fails to fulfill the Sakharov's conditions~\cite{Sakharov:1967dj} necessary for generating baryon asymmetry dynamically. In view of this, several beyond standard model (BSM) proposals have been put forward to solve these puzzles either one at a time or simultaneously. Among them, the weakly interacting massive particle (WIMP) paradigm of DM \cite{Kolb:1990vq, Jungman:1995df} and baryogenesis/leptogenesis \cite{Weinberg:1979bt, Kolb:1979qa, Fukugita:1986hr} have been the most widely studied ones. In WIMP framework, a particle having mass and interactions around the electroweak ballpark gives rise to the required DM relic after thermal freeze-out. On the other hand, typical baryogenesis scenarios involve out-of-equilibrium CP violating decay of heavy particles generating the observed matter-antimatter asymmetry. In leptogenesis~\cite{Fukugita:1986hr}, a non-zero asymmetry is first generated in the lepton sector which later gets converted into baryon asymmetry through $(B+L)$-violating electroweak (EW) sphaleron transitions~\cite{Kuzmin:1985mm}. One appealing feature of leptogenesis is the connection to light neutrino mass and mixing, another observed phenomena unexplained by the SM. The heavy degrees of freedom like right-handed neutrino (RHN) considered in leptogenesis can also give rise to light active neutrino masses via seesaw mechanism \cite{Minkowski:1977sc, GellMann:1980vs, Mohapatra:1979ia, Schechter:1980gr, Schechter:1981cv}.

While WIMP DM can have sizeable interactions with ordinary matter to be discovered at terrestrial experiments, null results at direct detection experiments \cite{LZ:2022lsv} have pushed WIMP DM parameter space to a tight corner. On the other hand, typical baryogenesis or leptogenesis remain a high scale phenomena keeping it out of reach from direct experimental reach. This has motivated alternative and indirect ways of probing such mechanisms providing interesting complementarities with usual laboratory experiments. One such avenue is the detection of stochastic gravitational wave (GW) background, which has been utilised in several baryogenesis or leptogenesis scenarios \cite{Hall:2019ank, Dror:2019syi, Blasi:2020wpy, Fornal:2020esl, Samanta:2020cdk, Barman:2022yos, Baldes:2021vyz, Azatov:2021irb, Huang:2022vkf, Dasgupta:2022isg, Barman:2022pdo, Datta:2022tab, Borah:2022cdx, Borah:2023saq, Borah:2023god, Barman:2023fad, Borah:2024qyo, Borah:2024bcr, Barman:2024ujh} as well as particle DM models \cite{Hall:2019rld, Yuan:2021ebu, Tsukada:2020lgt, Chatrchyan:2020pzh, Bian:2021vmi, Samanta:2021mdm, Borah:2022byb, Azatov:2021ifm, Azatov:2022tii, Baldes:2022oev, Borah:2022iym, Borah:2022vsu, Shibuya:2022xkj, Borah:2023saq, Borah:2023god, Borah:2023sbc, Borah:2024lml, Borah:2024qyo, Adhikary:2024btd, Borah:2024kfn, Borah:2024bcr, Barman:2024ujh, Borboruah:2024lli, Borah:2025wzl}. Motivated by this, here we consider a TeV scale leptogenesis scenario having gravitational wave signatures due to a strong first-order electroweak phase transition (EWPT). We consider a singlet RHN $N_1$ coupling to lepton and Higgs doublets in the SM which can also generate one light neutrino mass via type-I seesaw. A dark sector comprising of a chiral singlet fermion $\psi$, a scalar doublet $\eta$, a real scalar singlet $\chi$ all of which are odd under an unbroken $Z_2$ symmetry is considered which serve four important purposes: (i) providing one-loop contribution to $N_1$ decay into lepton and Higgs whose interference with the tree level decay provides non-zero CP asymmetry, (ii) generates another active neutrino mass at radiative level, (iii) provides a DM candidate in terms of the lightest $Z_2$-odd particle and (iv) turning the electroweak phase transition from a crossover to a strongly first-order phase transition (FOPT). The proposed setup is similar to the idea of the Higgs portal leptogenesis \cite{LeDall:2014too, Alanne:2018brf} where a type-I seesaw extended by a singlet scalar enhances the CP asymmetry due to the presence of additional parameters unrelated to the origin of neutrino mass. This leads to successful TeV scale leptogenesis without requiring any resonantly enhanced CP asymmetry \cite{Pilaftsis:2003gt}. While we have one additional field in the form of the $Z_2$-odd scalar doublet, it leads to additional phenomenology as mentioned above. Contrary to the pure type-I seesaw origin of light neutrino mass in Higgs portal leptogenesis works mentioned above, we have the popular {\it Scoto-Seesaw} scenario \cite{Rojas:2018wym} where the hierarchical atmospheric and solar neutrino mass scales can be generated from tree level and radiative contributions respectively. The DM phenomenology is similar to a WIMP setup with typical detection prospects at terrestrial experiments. The proposed scenario offers rich phenomenology due to the stochastic GW signatures from first-order EWPT with a variety of detection aspects at terrestrial experiments due to the possibility of all BSM particles to be within TeV scale.

This paper is organised as follows. In section \ref{sec1}, we discuss the details of the model followed by the details of leptogenesis and dark matter in section \ref{sec2}. We summarise the details of the first-order EWPT in section \ref{sec3} and discuss our numerical results in section \ref{sec4}. We finally conclude in section \ref{sec5}.

\section{The Model}
\label{sec1}
We consider a simple extension of the SM with two chiral singlet fermions $N_1, \psi$, one real singlet scalar $\chi$ and a scalar doublet $\eta$. 
We also incorporate an unbroken $Z_2$ symmetry under which $\psi, \eta, \chi$ are odd comprising the dark sector while all other fields are even. The lightest $Z_2$-odd particle thereby provides a suitable DM candidate. Table \ref{tab1} summarises the relevant particle content and the corresponding quantum numbers.

	\begin{table}[h]
    \centering
		\begin{tabular}{|c|cc|cccc|}
			\hline
			&$L$&$\Phi$& $N_1$ &$\psi$&$\eta$ & $\chi$\\
			\hline
			$SU(2)$&2&2&1&1&2 & 1\\
			$U(1)_Y$&$-\frac{1}{2}$&$\frac{1}{2}$&0&0& $\frac{1}{2}$ & 0\\
   		$Z_2$&$1$&$1$&1&-1&-1 & -1\\
			\hline
		\end{tabular}
		\caption{Particle content of the model with respective quantum numbers under the symmetry group.}\label{tab1}
	\end{table}

The relevant part of the Lagrangian is given by
\begin{align}
    -\mathcal{L}_Y \supset y_N \overline{\ell} \tilde{\Phi} N_1 + y_\psi \overline{\ell} \tilde{\eta} \psi + y_1 \overline{N^c_1} \psi \chi + \frac{1}{2} M_{N_1} \overline{N^c_1}N_1 + \frac{1}{2} M_\psi \overline{\psi^c} \psi + {\rm h.c.}
 \end{align}
The tree level scalar potential can be written as
\begin{align}
    V_{\rm tree} & =\mu_\Phi^2|\Phi|^2+\mu_\eta^2|\eta|^2+ \lambda_1|\Phi|^4 + \lambda_2|\eta|^4 +\lambda_3 |\Phi|^2|\eta|^2 +\lambda_4 |\eta^\dagger \Phi|^2 +\lambda_5[(\eta^\dagger \Phi)^2 +{\rm h.c.}] \nonumber \\
    & + \frac{\mu^2_\chi}{2} \chi^2 + \lambda_7 \chi^4 +\lambda_8 \chi^2 |\Phi |^2 + \lambda_9 \chi^2 |\eta|^2+ (\mu_1 \chi \Phi^\dagger \eta + \mu_1^* \chi \eta^\dagger \Phi).
    \label{treeV}
\end{align}
The unbroken $Z_2$ symmetry prevents $Z_2$-odd scalars $\eta, \chi$ from acquiring non-zero vacuum expectation value (VEV). The doublet scalar fields $\Phi$ and $\eta$ are parameterized as
\begin{align}
    \Phi=\frac{1}{\sqrt{2}} \begin{pmatrix}
        0 \\
        \phi +v
    \end{pmatrix}, \eta= \begin{pmatrix}
        \eta^\pm \\
        \frac{(H +i A)}{\sqrt{2}}
    \end{pmatrix}.
\end{align}
After electroweak symmetry breaking, neutrinos acquire a Dirac mass term $M_D= \frac{1}{\sqrt{2}} v y_N$. The tree level contribution to light neutrino mass in the seesaw limit $M_D \ll M_{N_1}$ is 
\begin{equation}
    M^{\rm tree}_\nu = -M_D M^{-1}_{N_1} M^T_D = -\frac{v^2}{2} \frac{y^{\alpha 1}_N y^{\beta 1}_{N}}{M_{N_1}}.
    \end{equation}
Since there is only one RHN taking part in tree level seesaw, it generates one of the light neutrino masses. On the other hand, the $Z_2$-odd singlet fermion $\psi$ and scalar doublet $\eta$ give rise to one-loop contribution to light neutrino mass in scotogenic fashion \cite{Tao:1996vb, Ma:2006km}
\begin{align}
M^{\rm loop}_{\nu} &=  \frac{y^{\alpha 1}_\psi y^{\beta 1}_\psi M_{\psi}}{32 \pi^2} \bigg ( \cos^2{\theta} F_1 (\eta_1) + \sin^2{\theta} F_1 (\eta_2) - F_1 (A) \bigg ) 
\label{numass1}
\end{align}
where $F_1 (x)=\frac{m^2_{x}}{m^2_{x}-M^2_\psi} \: \text{ln} \frac{m^2_{x}}{M^2_\psi} $, $\eta_{1,2}$ are the physical scalars resulting from diagonalising $2\times2$ scalar mass matrix in $(H, \chi)$ basis with a rotation matrix of angle $\theta$. The details of the scalar sector of the model can be found in \cite{Beniwal:2020hjc}. In the spirit of scoto-seesaw model, one can identify $M^{\rm tree}_\nu \sim \sqrt{\Delta m^2_{\rm atm}}, M^{\rm loop}_{\nu} \sim \sqrt{\Delta m^2_{\rm sol}}$ explaining the relative hierarchy between solar and atmospheric neutrino mass splitting from the loop suppression. In order to incorporate the constraints from light neutrino masses, we use the 
Casas-Ibarra (CI) parametrisation \cite{Casas:2001sr} for type-I seesaw in combination with the one for scotogenic model \cite{Toma:2013zsa}, as done for scoto-seesaw scenarions in \cite{Leite:2023gzl}. This is given by 
\begin{equation}
    \mathcal{Y} = \sqrt{X^{-1}_M} R \sqrt{m^{\rm diag}_\nu} U^\dagger
\end{equation}
where $R$ is an arbitrary complex orthogonal matrix satisfying $RR^{T}=\mathbbm{1}$ and $U$ is the usual Pontecorvo-Maki-Nakagawa-Sakata (PMNS) mixing matrix which diagonalises the light neutrino mass matrix in a basis where charged lepton mass matrix is diagonal. The combined Dirac Yukawa coupling $\mathcal{Y}$ is 
\begin{equation}
    \mathcal{Y} = \begin{pmatrix}
        y^{1\times 3}_{\psi} \\
    y^{1 \times 3}_N
    \end{pmatrix}
    \label{eq:yuk}
\end{equation}
and $X_M$ is given by
\begin{equation}
   X_M= \begin{pmatrix}
        -X_1 & 0 \\
        0 & X_0
    \end{pmatrix}, \,\, X_1 = \frac{M_{\psi}}{32 \pi^2} \bigg ( \cos^2{\theta} F_1 (\eta_1) + \sin^2{\theta} F_1 (\eta_2) - F_1 (A) \bigg ), \,\, X_0=\frac{v^2}{2M_{N_1}}.
\end{equation}
The $R$ matrix for 2 heavy neutrino scenario is given by \cite{Ibarra:2003up}
\begin{equation}
    R = \begin{pmatrix}
        0 & \cos{z} & \pm \sin{z} \\
        0 & -\sin{z} & \pm \cos{z}
    \end{pmatrix}
\end{equation}
where $z=a+ib$ is a complex angle. The diagonal light neutrino mass matrix, assuming normal hierarchy (NH), is given by 
\begin{equation}
    m^{\rm diag}_\nu = \begin{pmatrix}
        0 & 0 & 0 \\
        0 & \sqrt{\Delta m^2_{\rm sol}} & 0 \\
        0 & 0 & \sqrt{\Delta m^2_{\rm atm}}
    \end{pmatrix}.
\end{equation}

\section{Leptogenesis and dark matter}
\label{sec2}
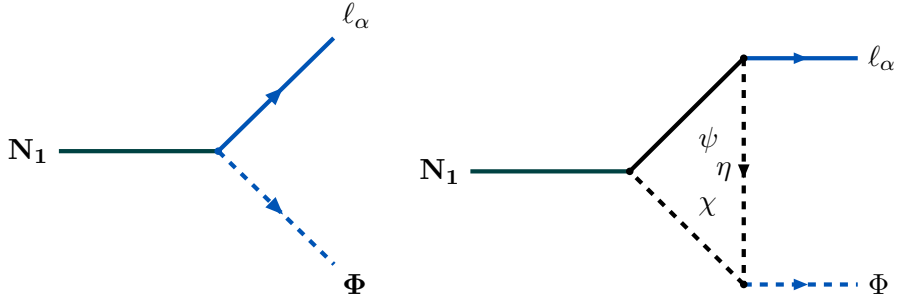
\begin{figure}[htb!]
\centering
{\begin{tikzpicture}
\begin{feynman}
\vertex (a){\(\rm\bf\color{black}{N_1 }\)};
\vertex [right=2.5cm of a] (b);
\vertex [above right=1.5cm and 1.5cm of b] (c){\(\rm\bf\color{black}{\ell_{\alpha} }\)};
\vertex [below right=1.5cm and 1.5cm of b] (d){\(\rm\bf\color{black}{\Phi }\)};
\diagram* {
(a) -- [line width=0.25mm,plain,  style=warmblack,ultra thick] (b),
(b)-- [line width=0.25mm,fermion,  style=mediumtealblue,ultra thick] (c),
(b) -- [line width=0.25mm,charged scalar, style=mediumtealblue,ultra thick] (d)};
\node at (b)[circle,fill,style=mediumtealblue,inner sep=1pt]{};
\end{feynman}
\end{tikzpicture}}\quad
{\begin{tikzpicture}
\begin{feynman}
\vertex (a){\(\rm\bf\color{black}{N_1 }\)};
\vertex [right=2.5cm of a] (b);
\vertex [above right=1.5cm and 1.5cm of b] (c);
\vertex [below right=1.5cm and 1.5cm of b] (d);
\vertex [right=1.5cm of c] (e){\color{black}{\(\ell_{\alpha}\)}};
\vertex [right=1.5cm of d] (f){\color{black}{\(\Phi\)}};
\diagram* {(a) -- [line width=0.25mm,plain,arrow size=1.2pt, style=warmblack,ultra thick] (b),
(b)-- [line width=0.25mm, plain, arrow size=1.2pt, style=black,ultra thick, edge label'={\(\rm\bf\color{black}{\psi }\)}] (c),
(d) -- [line width=0.25mm,scalar, style=black, ultra thick,arrow size=1.2pt,edge label'={\(\rm\color{black}{\chi }\)} ] (b),
(c) -- [line width=0.25mm,charged scalar, style=black, ultra thick, arrow size=1.2pt, edge label'={\(\rm\bf\color{black}{\eta}\)}] (d),
(c) -- [line width=0.25mm,fermion,  arrow size=1.2pt, style=mediumtealblue,ultra thick] (e),
(d) -- [line width=0.25mm,charged scalar, style=mediumtealblue, ultra thick,arrow size=1.2pt ] (f)};
\node at (b)[circle,fill,style=black,inner sep=1pt]{};
\node at (c)[circle,fill,style=black,inner sep=1pt]{};
\node at (d)[circle,fill,style=black,inner sep=1pt]{};
\end{feynman}
\end{tikzpicture}}
\caption{Tree level and one-loop Feynman diagrams for dark sector portal leptogenesis.}
\label{fig:feynman-lepto}
\end{figure}

Assuming $M_{N_1} < M_\psi$, lepton asymmetry generated by out-of-equilibrium decay of $N_1$ at a lower scale survives while asymmetries generated by $\psi$ at a higher scale gets washed out. The non-zero CP asymmetry in the decay of $N_1 \rightarrow \ell \Phi$ arises from the interference of tree level and one-loop vertex correction shown in Fig. \ref{fig:feynman-lepto}. Unlike in the original Higgs portal leptogenesis with type-I seesaw \cite{LeDall:2014too, Alanne:2018brf}, there is no self-energy contribution to the CP asymmetry in our setup ruling out the possibility of resonant enhancement. However, due to the presence of new parameters $y_1, \mu_1$ relating dark sector particles with others, it is possible to enhance the CP asymmetry even for TeV scale $N_1$, while being consistent with light neutrino mass. Due to the chosen hierarchy $M_{N_1} < M_\psi$, we consider $Z_2$-odd scalars to be lighter than $M_{N_1}$ such that an imaginary part survives from the one-loop diagram leading to non-vanishing CP asymmetry. This also ensures that DM (lightest physical state among $\chi, H, A$) remains in equilibrium during the generation of lepton asymmetry such that its freeze-out can be studied independently.

With these assumptions, the CP asymmetry can be found as \cite{LeDall:2014too, Alanne:2018brf, Bhattacharya:2024ohh}
\begin{align}
    \epsilon_1 &= \frac{\Gamma (N_1 \rightarrow \ell \Phi)-\Gamma (N_1 \rightarrow \overline{\ell} \Phi^\dagger)}{\Gamma (N_1 \rightarrow \ell \Phi)+\Gamma (N_1 \rightarrow \overline{\ell} \Phi^\dagger)} = \frac{\Gamma (N_1 \rightarrow \ell \Phi)-\Gamma (N_1 \rightarrow \overline{\ell} \Phi^\dagger)}{\Gamma_N} \nonumber \\
     &= \frac{1}{8 \pi}\frac{{\rm Im}(y_N^\dagger y_\psi y_1 \mu_1)}{(y_N^\dagger y_N)M_{N_1} (1-\xi+\omega)\sqrt{(1-\xi+\omega)^2-4\omega}} \bigg ( 1+\xi -2\sqrt{\sigma} \nonumber \\
    & +(\delta-\sqrt{\delta}(1-\xi+\omega)-\zeta+\omega) {\rm ln}\left[ \frac{\delta-(1-\sqrt{\sigma})^2}{\delta-\sigma+\xi}\right] \bigg ),
\end{align}
where $\delta =\frac{M_\psi^2}{M_{N_1}^2}$, $\zeta=\frac{m_\eta^2}{M_{N_1}^2}$, $\xi=\frac{m_h^2}{M_{N_1}^2}$, $\omega=\frac{m_l^2}{M_{N_1}^2}$ and $\sigma=\frac{m_\chi^2}{M_{N_1}^2}$. In the limit of vanishing SM Higgs and lepton mass, the CP asymmetry simplifies to
\begin{align}
    \epsilon_1=\frac{1}{8 \pi}\frac{{\rm Im}(y_N^\dagger y_\psi y_1 \mu_1)}{(y_N^\dagger y_N)M_{N_1}} \left(1-2\sqrt{\sigma}+(\delta-\sqrt{\delta}-\zeta){\rm ln}\left[ \frac{\delta-(1-\sqrt{\sigma})^2}{\delta-\sigma}\right] \right).
\end{align}
The corresponding Boltzmann equations for comoving densities of $N_1$ and $B-L$ can be written as
\begin{align}
    \frac{dY_{\rm N_1}}{dz} =  & -D_{\rm N}\left(Y_{\rm N_1}-Y_{\rm N_1}^{\rm eq}\right) - \frac{s}{\rm H(z)z} \left(Y_{\rm N_1}^{2}-\left( Y_{\rm N_1}^{\rm eq} \right)^{2}\right)  \bigg [\langle \sigma v \rangle_{N_1 N_1 \longrightarrow \chi \chi} + \langle \sigma v \rangle_{N_1 N_1 \longrightarrow \Phi \Phi^{\dagger}} \nonumber \\
    & + \langle \sigma v \rangle_{N_1 N_1 \longrightarrow \ell_{\alpha} \overline{\ell}_{\beta}}\bigg ]- \frac{s}{\rm H(z)z} \left(Y_{\rm N_1}-Y_{\rm N_1}^{\rm eq}\right) \bigg [2Y_{l}^{\rm eq}\langle \sigma v \rangle_{\overline{l} N_1 \longrightarrow \overline{q}t} + 4Y_{t}^{\rm eq}\langle \sigma v \rangle_{ N_1 t \longrightarrow \overline{l}q} \nonumber 
      \\
     & + 2Y_{\chi}^{\rm eq}\langle \sigma v \rangle_{N_1 \chi \longrightarrow \overline{\ell} \eta^{\dagger}} + 2Y_{\Phi}^{\rm eq}\langle \sigma v \rangle_{N_1 \Phi \longrightarrow l_{\alpha} V_{\mu}}+ 2y_{\psi}^{\rm eq}\langle \sigma v \rangle_{N_1 \psi \longrightarrow \Phi \eta^{\dagger}}\bigg ],
\end{align}
\begin{align}
    \frac{dY_{\rm B-L}^{}}{dz} = & - \epsilon_{1}D_{\rm N}\left(Y_{\rm N_1}-Y_{\rm N_1}^{\rm eq}\right) - W_{\rm ID} Y_{\rm B-L}
     - \frac{\rm s}{\rm H(z)z} Y_{\rm B-L} \Big[ 2Y_{\rm \Phi}^{\rm eq} \langle \sigma v \rangle_{l \Phi^{\dagger}\longrightarrow \overline{\ell} \Phi} 
     +Y_{\rm N_1} \langle \sigma v \rangle_{\overline{\ell}N_1 \longrightarrow \overline{q}t} \nonumber \\
     & + 2 Y_q^{\rm eq} \langle \sigma v \rangle_{\overline{\ell}q \longrightarrow N_1 t} + 2 Y_{l}^{\rm eq} \langle \sigma v  \rangle_{\ell \ell \longrightarrow \Phi^{\dagger}\Phi^{\dagger}} + Y_{\eta}^{\rm eq} \langle \sigma v  \rangle_{\overline{\ell} \eta^{\dagger} \longrightarrow N_1 \chi } + Y_{V}^{\rm eq} \langle \sigma v \rangle_{\overline{\ell} V_{\mu}\longrightarrow \Phi N_1} \Big].
\end{align}
Here $Y_i=n_i/s$ denotes comoving density with $n_i$ being number density of species 'i' and $s=\frac{2\pi^2}{45}g_{*s} T^3$ being entropy density of the Universe. $z=M_{N_1}/T$ and $H(z) =\sqrt{\frac{4\pi^3 g_*(T)}{45}}\frac{T^2}{M_{\rm Pl}}$ is the Hubble parameter at high temperatures where $g_{*s}$ remains constant. The decay term $D_N$ is defined as $D_N  =  \dfrac{ \langle \Gamma_{N} \rangle}{{H} z} = K_{N}z\dfrac{\kappa_{1}(z)}{\kappa_{2}(z)}$ where $K_N=\Gamma_N/{ H}(z=1)$ with $\kappa_i(z)$ being the modified Bessel function of $i$-th kind. The washout due to inverse decay is $W_{\rm ID} =  \dfrac{1}{4}K_{N}z^{3}\kappa_{1}(z)$. The thermal averaged cross-section is defined as \cite{Gondolo:1990dk}
\begin{equation}
    \langle \sigma v \rangle_{ij \rightarrow kl} = \frac{1}{8Tm^2_i m^2_j \kappa_2 (z_i) \kappa_2 (z_j)} \int^{\infty}_{(m_i+m_j)^2} ds \frac{\lambda (s, m^2_i, m^2_j)}{\sqrt{s}} \kappa_1 (\sqrt{s}/T) \sigma
\end{equation}
with $z_i=m_i/T$ and $\lambda (s, m^2_i, m^2_j)=[s-(m_i+m_j)^2][s-(m_i-m_j)^2]$.

The final baryon asymmetry $\eta_B$ can be analytically estimated to be \cite{Buchmuller:2004nz}
 \begin{align}
     \eta_B = \frac{a_{\rm sph}}{f} \epsilon_1 \kappa\,, \label{eqn:etaana}
 \end{align}
where the factor $f$ accounts for the change in the relativistic degrees of freedom from the scale of leptogenesis until recombination and comes out to be $f=\frac{106.75}{3.91}\simeq27.3$. $\kappa$ is known as the efficiency factor which incorporates the effects of washout processes while $a_\text{Sph}$ is the sphaleron conversion factor. The lepton asymmetry at the sphaleron decoupling epoch $T_{\rm sph} \sim 130$ GeV gets converted into baryon asymmetry as \cite{Harvey:1990qw}
\begin{align}
& Y_B\simeq a_\text{Sph}\,Y_{B-L}=\frac{8\,N_F+4\,N_H}{22\,N_F+13\,N_H}\,Y_{B-L} =\frac{8}{23} Y_{B-L}\,, 
\label{eq:sphaleron}
\end{align}
with $N_F=3\,,N_H=2$ being the fermion generations and the number of scalar doublets in our model respectively. The observational constraint on $\eta_B$ given in Eq. \eqref{etaBobs} can be translated to $Y_B$ as
\begin{equation}
    Y_B (T_0) = 8.67 \times 10^{-11}.
\end{equation}

Dark matter relic can similarly be determined by solving the corresponding Boltzmann equation
\begin{equation}
     \frac{dY_{\rm DM}}{dx} = -\frac{s}{H(x)x}(Y_{\rm DM}^{2}-(Y_{\rm DM}^{\rm eq})^{2})\langle\sigma v \rangle_{\rm DM \, DM \rightarrow SM \, SM}
\end{equation}
where $x=m_{\rm DM}/T$ and $\langle\sigma v \rangle_{\rm DM \, DM \rightarrow SM \, SM}$ denotes the thermal averaged annihilation cross section of DM into SM particles. As mentioned earlier, depending upon the parameter space, one of the $Z_2$-odd neutral scalars $\eta_1, \eta_2, A$ play the role of DM. 
%{\small
%\begin{eqnarray}\begin{split}
%\dfrac{dY_{\rm N}}{dz}  = &-D_{\rm N}\left(Y_{\rm N}-Y_{\rm N}^{eq}\right)-\dfrac{s}{\rm H(z)z} \left(Y_{\rm N}^{2}-\left( Y_{\rm N}^{eq} \right)^{2}\right)\left[\langle \sigma v \rangle_{NN\longrightarrow \chi \chi}+\langle \sigma v \rangle_{NN\longrightarrow \Phi \Phi^{\dagger}}+\langle \sigma v \rangle_{NN\longrightarrow \ell_{\alpha} \overline{\ell}_{\beta}}\right]\,, \\
%&-\dfrac{\rm s}{\rm H(z)z} \left(Y_{\rm N}-Y_{\rm N}^{eq}\right)\left[Y_{\Phi}^{eq}\langle \sigma v  \rangle_{N\Phi \longrightarrow l \eta^{\dagger}}+Y_{\Phi}^{eq}\langle \sigma v \rangle_{N\Phi\longrightarrow l_{\alpha} V_{\mu}}+y_{\psi}^{eq}\langle \sigma v \rangle_{N\psi\longrightarrow \Phi \Phi^{\dagger}}\right]\\
%\dfrac{dY_{\rm \Delta L}^{}}{dz} = & \epsilon_{\rm N}^{}D_{\rm N}\left(Y_{\rm N}-Y_{\rm N}^{eq}\right)-W_{\rm ID} Y_{\rm \Delta L}-\dfrac{\rm s}{\rm H(z)z} Y_{\rm \Delta L}\left[ Y_{\rm \Phi^{}}^{eq}  \langle \sigma v \rangle_{l \Phi^{\dagger}\longrightarrow \overline{l} \Phi} \right.\\ &\left. + Y_{ l}^{eq}r^2_{\Phi} \langle \sigma v  \rangle_{\Phi^{\dagger}\Phi^{\dagger} \longrightarrow l l}+ \frac{1}{2}Y_{l}^{eq} r_{\rm N} r_{\rm \chi} \langle \sigma v  \rangle_{N \chi \longrightarrow \overline{l} \eta^{\dagger}} + \frac{1}{2}Y_{l}^{eq} r_{\rm N}r_{\Phi} \langle \sigma v \rangle_{\Phi N\longrightarrow \overline{l} V_{\mu}} +\frac{1}{2} Y_{l}^{eq}r_{N}r_{t} \langle \sigma v  \rangle_{N t\longrightarrow l t} \right].
%\label{eq:cbeq-lepto}
%\end{split}\end{eqnarray}}

%\noindent Here $r_{i}=Y_{i}^{eq}/Y_{l}^{eq}$.

\section{First-order electroweak phase transition}
\label{sec3}
In order to study the high temperature behaviour of the scalar potential, we first calculate the complete potential including the tree level potential $V_{\rm tree}$, one-loop Coleman-Weinberg potential $V_{\rm CW}$\cite{Coleman:1973jx} along with the finite-temperature potential $V_{\rm th}$ \cite{Dolan:1973qd,Quiros:1999jp}. Some recent reviews on first-order phase transition can be found in \cite{Mazumdar:2018dfl,Hindmarsh:2020hop, Athron:2023xlk}.

The corresponding effective potential can be written as
\begin{equation}
    V_{\rm eff}= V_{\rm tree}+V_{\rm CW}+V_{\rm th} + V_{\rm daisy}.
    \label{eq:Veff}
\end{equation}
While the $V_{\rm tree}$ is given by Eq. \eqref{treeV}, the Coleman-Weinberg potential~\cite{Coleman:1973jx} with $\overline{\rm DR}$ regularisation is given by
\begin{align}
V_{\rm CW} = \sum_i (-)^{n_{f}} \frac{n_i}{64\pi^2} m_i^4 (\phi) \left(\log\left(\frac{m_i^2 (\phi)}{\mu^2} \right)-C_i \right),
\end{align}
where suffix $i$ represents particle species, and $n_i,~m_i (\phi)$ are the degrees of freedom (dof) and field-dependent masses of $i$'th particle, written as a function of the neutral component of the SM Higgs field $\phi$, details of which are given in appendix \ref{appen1}.
In addition, $\mu$ is the renormalisation scale, and $(-)^{n_f}$ is $+1$ for bosons and $-1$ for fermions, respectively. The thermal contributions to the effective potential can be written as
\begin{align}
V_{\rm th} = \sum_i \left(\frac{n_{\rm B_i}}{2\pi^2}T^4 J_B \left[\frac{m_{\rm B_i}}{T}\right] - \frac{n_{\rm F_{i}}}{2\pi^2}T^4 J_F \left[\frac{m_{\rm F_{i}}}{T}\right]\right),
\end{align}
where $n_{B_i}$ and $n_{F_i}$ denote the dof of the bosonic and fermionic particles, respectively and  $J_B, J_F$ are defined by following functions:
\begin{align}
J_B(x) =\int^\infty_0 dz z^2 \log\left[1-e^{-\sqrt{z^2+x^2}}\right], \,
J_F(x) =   \int^\infty_0 dz z^2 \log\left[1+e^{-\sqrt{z^2+x^2}}\right]. \label{eq:J_B} 
\end{align}
We also include the Daisy corrections \cite{Fendley:1987ef,Parwani:1991gq,Arnold:1992rz} which improve the perturbative expansion during the FOPT. While there are two schemes namely, Parwani method and Arnold-Espinosa method, we use the latter. The Daisy contribution, in this scheme, is given by
\begin{equation}
        V_{\rm daisy}(\phi,T) = -\sum_i \frac{g_i T}{12\pi}\left[ m^3_i(\phi,T) - m^3_i(\phi) \right]
\end{equation}
The thermal masses for different components are given by $m^2_i(\phi,T)=m^2_i(\phi) + \Pi_i(T)$ the details of which are given in appendix \ref{appen1}.

We consider a single step FOPT where only the neutral component of the SM Higgs doublet $\phi$ acquires a non-zero VEV. Using the full finite-temperature potential, we then calculate the critical temperature $T_c$ at which the scalar potential develops a second degenerate minima at $v_c = \phi (T=T_c)$. This also decides the order parameter of the FOPT defined as $v_c/T_c$, a larger value of which implies a stronger phase transition. Once the second minima appears, the FOPT proceeds via tunneling of the false vacuum $(\phi=0)$ to the true vacuum $(\phi \neq 0)$. The rate of tunneling is estimated by calculating the bounce action $S_3$ using the prescription in \cite{Linde:1980tt, Adams:1993zs}. The nucleation temperature $T_n$ is then calculated by comparing the tunneling rate $\Gamma$ with the Hubble expansion rate of the universe $\Gamma (T_n) = H^4(T_n) \equiv H^4_*$.

One of the most interesting features of strong FOPT in the early Universe is the generation of stochastic gravitational wave background due to bubble collisions~\cite{Turner:1990rc,Kosowsky:1991ua,Kosowsky:1992rz,Kosowsky:1992vn,Turner:1992tz}, the sound wave of the plasma~\cite{Hindmarsh:2013xza,Giblin:2014qia,Hindmarsh:2015qta,Hindmarsh:2017gnf} and the turbulence of the plasma~\cite{Kamionkowski:1993fg,Kosowsky:2001xp,Caprini:2006jb,Gogoberidze:2007an,Caprini:2009yp,Niksa:2018ofa}. The total GW spectrum is then given by 
$$\Omega_{\rm GW}(f) = \Omega_\phi(f) + \Omega_{\rm sw}(f) + \Omega_{\rm turb}(f).$$ 
While the peak frequency and peak amplitude of such GW spectrum depend upon specific FOPT related parameters, the exact nature of the spectrum is determined by numerical simulations. The details of the GW spectrum from all three sources mentioned above are given in appendix \ref{appen2}. The key parameters relevant for GW estimates namely, the inverse duration of the phase transition and the latent heat released are calculated and parametrised in terms of \cite{Caprini:2015zlo}
$$\frac{\beta}{{ H}(T)} \simeq T\frac{d}{dT} \left(\frac{S_3}{T} \right) $$ and 
$$ \alpha_* =\frac{1}{\rho_{\rm rad}}\left[\Delta V_{\rm tot} - \frac{T}{4} \frac{\partial \Delta V_{\rm tot}}{\partial T}\right]_{T=T_n} $$ 
respectively, where $\Delta V_{\rm tot}$ is the energy difference in true and false vacua. The bubble wall velocity $v_w$ is estimated from the Jouguet velocity \cite{Kamionkowski:1993fg, Steinhardt:1981ct, Espinosa:2010hh}
$$v_J = \frac{1/\sqrt{3} + \sqrt{\alpha^2_* + 2\alpha_*/3}}{1+\alpha_*}$$
following the prescription given in \cite{Lewicki:2021pgr}. While the release of vacuum energy can reheat the Universe briefly potentially diluting the lepton asymmetry generated at a higher scale, such entropy dilution is negligible in our scenario as we never enter the supercooled regime of FOPT.

%\begin{figure}
%    \centering
%    \includegraphics[width=0.6\linewidth]{Plot1.pdf}
%    \caption{Evolution of asymmetry: $M_N=1058$ GeV, $M_\psi=10^4$ GeV, $y_1=10^{-3}$, $\mu_1=320i$ GeV, and other parameters are same BP1 of Table \ref{tab1}.}
%    \label{fig:enter-label}
%\end{figure}

\begin{figure}
    \centering
    \includegraphics[width=0.48\linewidth]{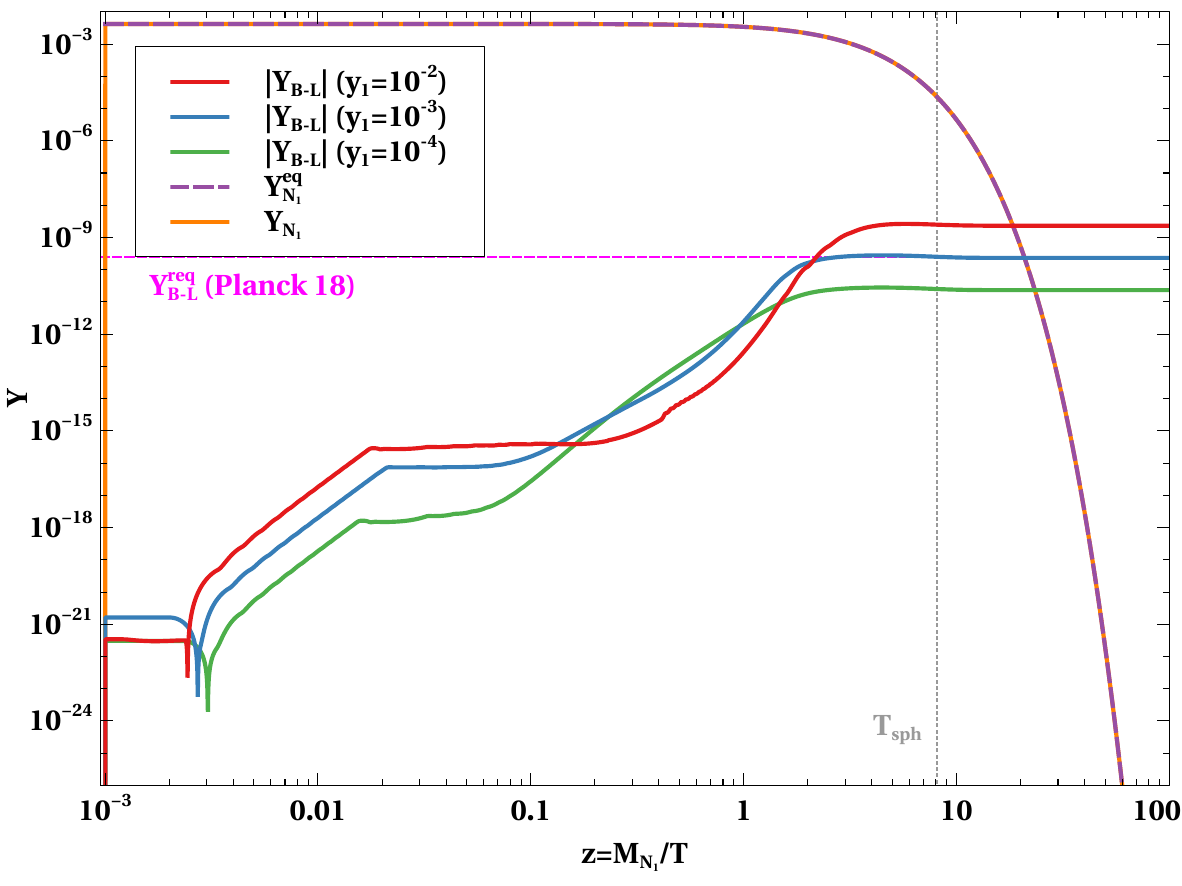}
    \includegraphics[width=0.48\linewidth]{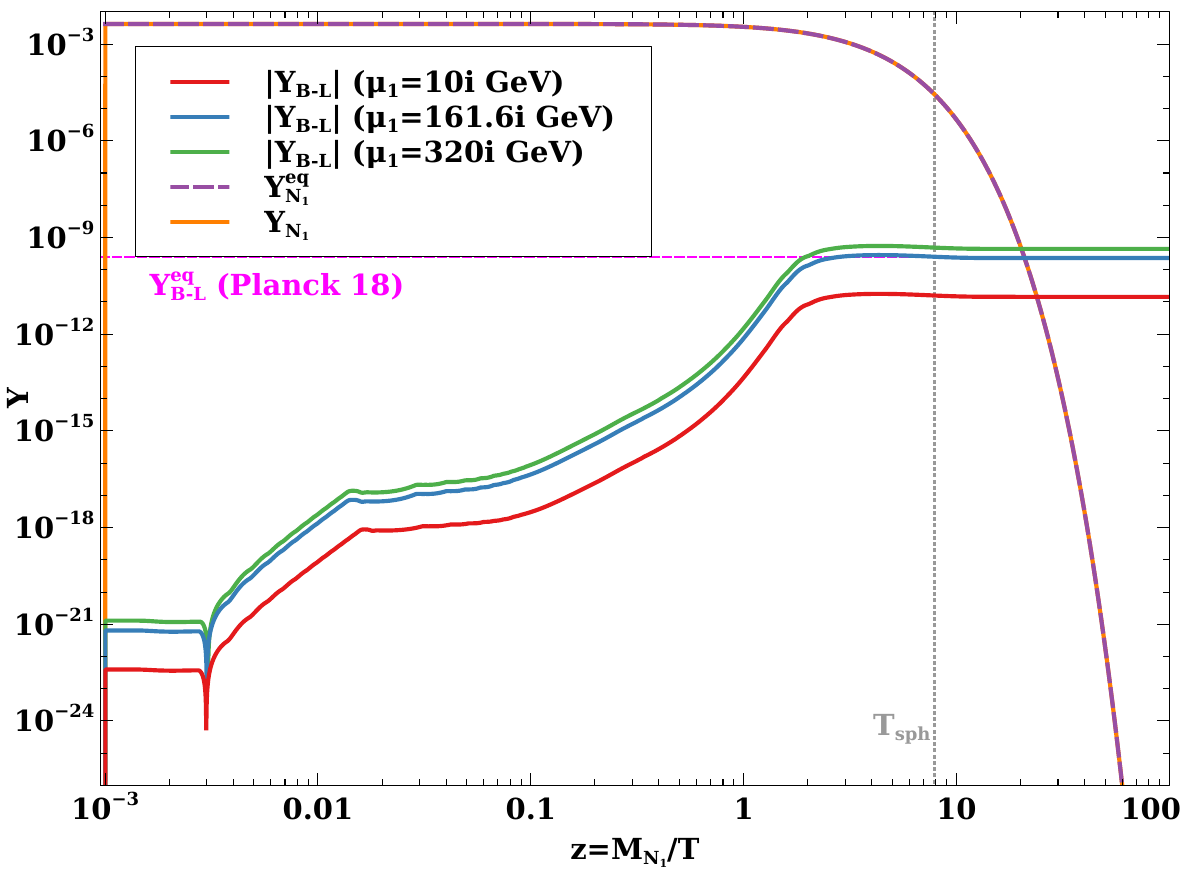}
    \caption{Left panel: evolution of lepton asymmetry for $M_{N_1}=1058$ GeV, $M_\psi=10^4$ GeV, $\mu_1=320i$ GeV and varying $y_1$ with other parameters being same as BP1 of table \ref{tab1}. Right panel: same as left panel but for different values of $\mu_1$. In both the panels, the vertical dotted line indicates the sphaleron decoupling epoch $T_{\rm sph} \sim 130$ GeV.}
    \label{fig:lepto}
\end{figure}

\begin{figure}
    \centering
   \includegraphics[width=0.5\linewidth]{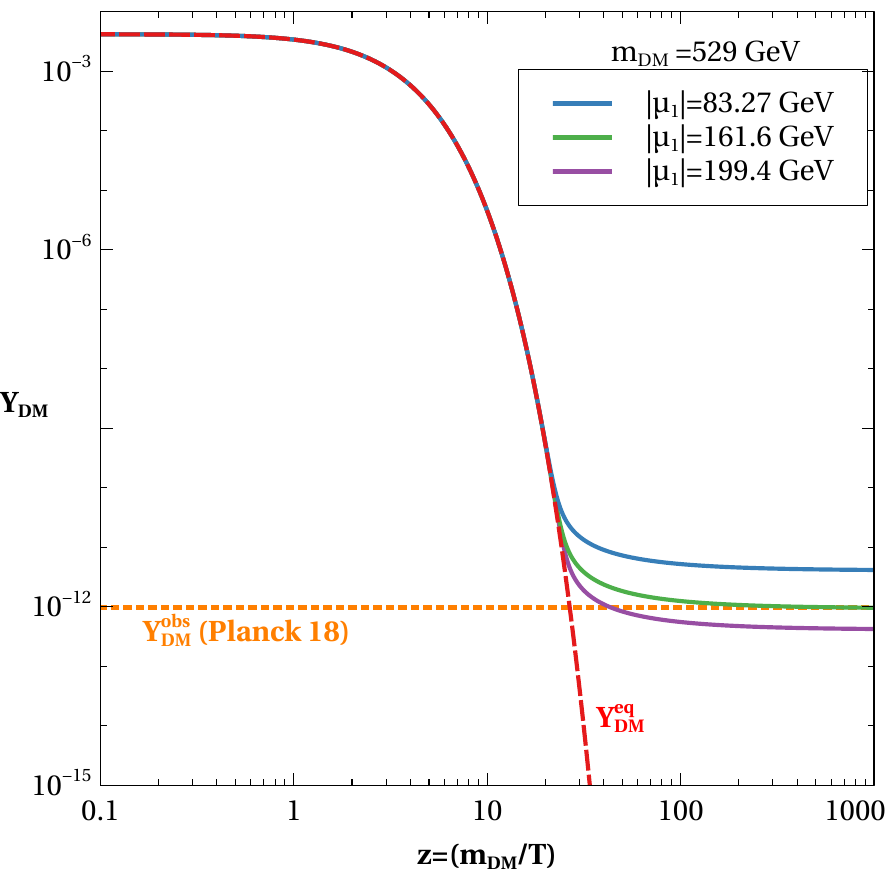}
    \caption{Dark matter evolution corresponding to BP1 of table \ref{tab1} for different values of $\mu_1$.}
    \label{fig:DM}
\end{figure}

%\begin{figure}
%    \centering
%    \includegraphics[width=0.6\linewidth]{Plot4.pdf}
%    \caption{Evolution of asymmetry: $M_N=2000$ GeV, $M_\psi=10^4$ GeV, $y_1=10^{-3}$, $\mu_1=18.5i$ GeV, $\mu_\chi=430$ GeV, $\mu_\eta=450$ GeV, $M_H=529$ GeV, $M_A =M_\eta=682$ GeV, $M_\chi=600$ GeV}
%    \label{fig:enter-label}
%\end{figure}

\begin{figure}
    \centering
      \includegraphics[width=0.48\linewidth]{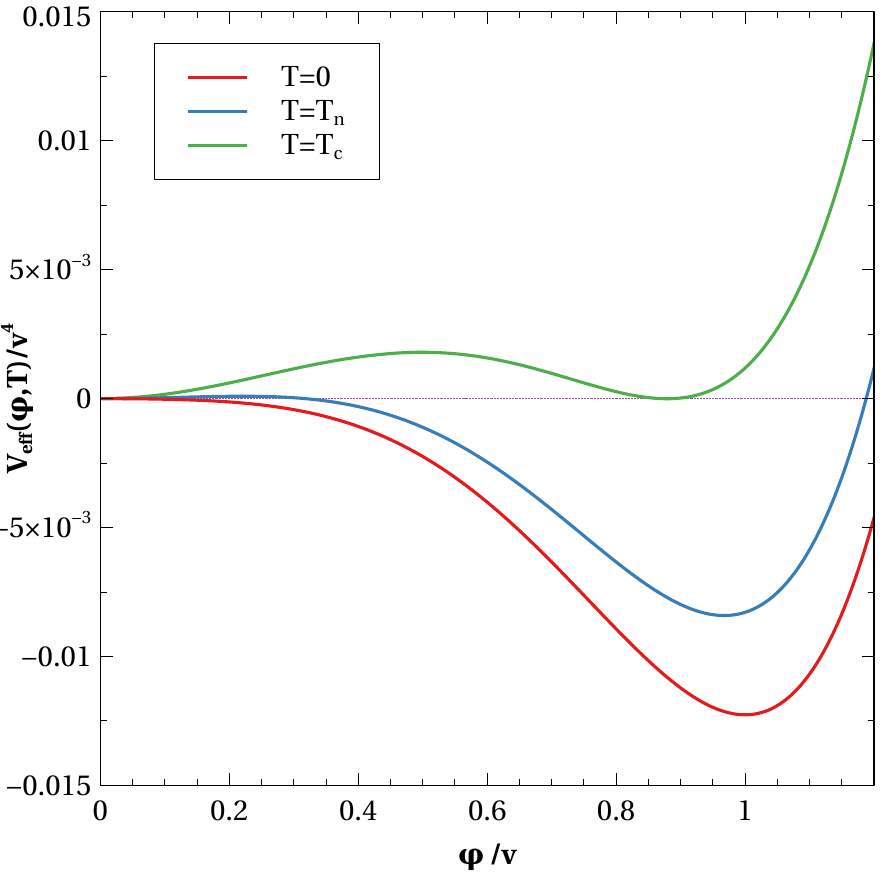}
    \includegraphics[width=0.48\linewidth]{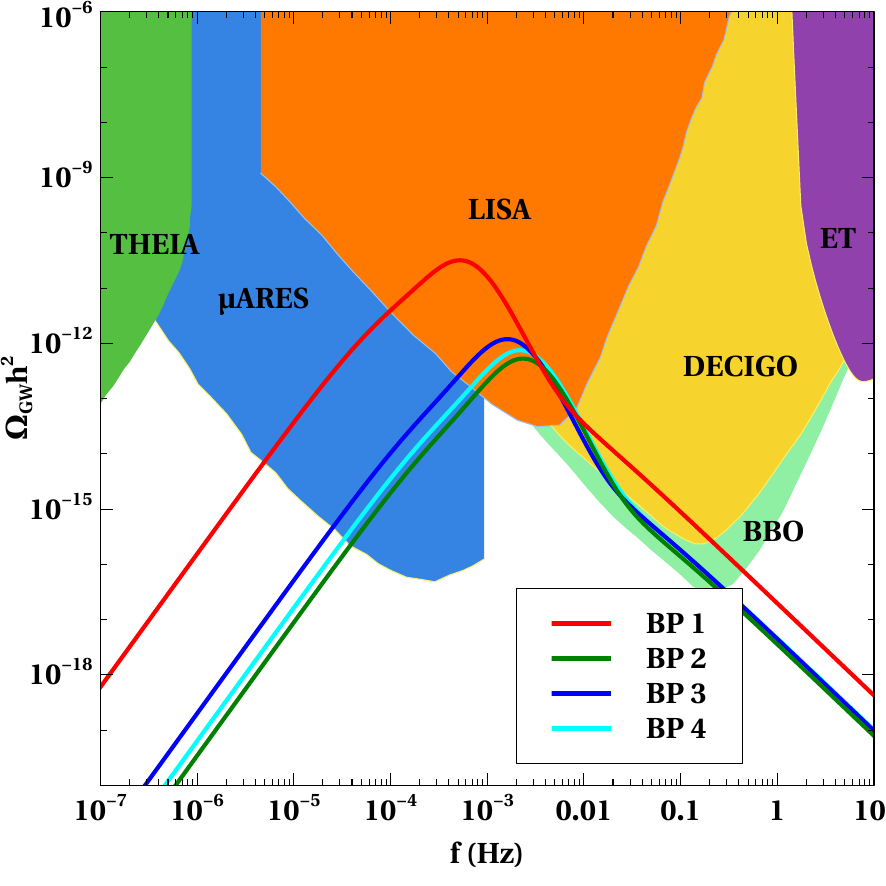}
    \caption{Left panel: profile of the effective potential corresponding to BP1 of table \ref{tab1} for different temperatures $T=0, T_n, T_c$. Right panel: GW spectrum corresponding to the benchmark points given in table \ref{tab1}.}
    \label{fig:gw}
\end{figure}

 \begin{table}[]
 \footnotesize
    \centering
    \begin{tabular}{|c|c|c|c|c|c|c|c|c|c|c|c|c|}
    \hline
         &  $\mu_\eta$ & $\mu_\chi$ & $M_{\eta_1}$ & $M_\eta (M_A)$ & $M_{\eta_2}$ & $\lvert \mu_1 \rvert$ & $y_1$ & $T_c$ & $T_n$ & $\beta/H_*$ & $\alpha_*$ & $M_{N_1}$\\
          & (GeV) & (GeV) & (GeV) & (GeV) & (GeV) & (GeV) & $(10^{-3})$ & (GeV) & (GeV) & & & (TeV) \\
         \hline
         BP1 & 450 & 430 & 529 & 682 & 600 & 161.6 &$2$ & 65.6 & 39.1 & 136.0 & 0.52 & 1.02\\
         \hline
         BP2 & 938 & 592 & 1046 & 1091 & 815 & 733.5 & $1$ & 66.5 & 48.9 &443.6 & 0.20 & 1.64 \\
         \hline
         BP3 & 660 & 220 & 714 & 778 & 598 & 301.2 & $1$ & 71.6 & 48.8 & 308.2 & 0.21 & 0.95\\
         \hline 
         BP4 & 524 & 57 & 820 & 648 & 336 & 461.1 & $1$ & 67.2 & 47.9 & 429.4 & 0.23 & 2.54\\
         \hline
    \end{tabular}
    \caption{Benchmark points satisfying the requirements of observed baryon asymmetry and dark matter. The mass of $Z_2$-odd heavy fermion is kept at $M_\psi \sim 10 M_{N_1}$.}
    \label{tab1}
\end{table}

\begin{figure}
    \centering
   \includegraphics[width=0.48\linewidth]{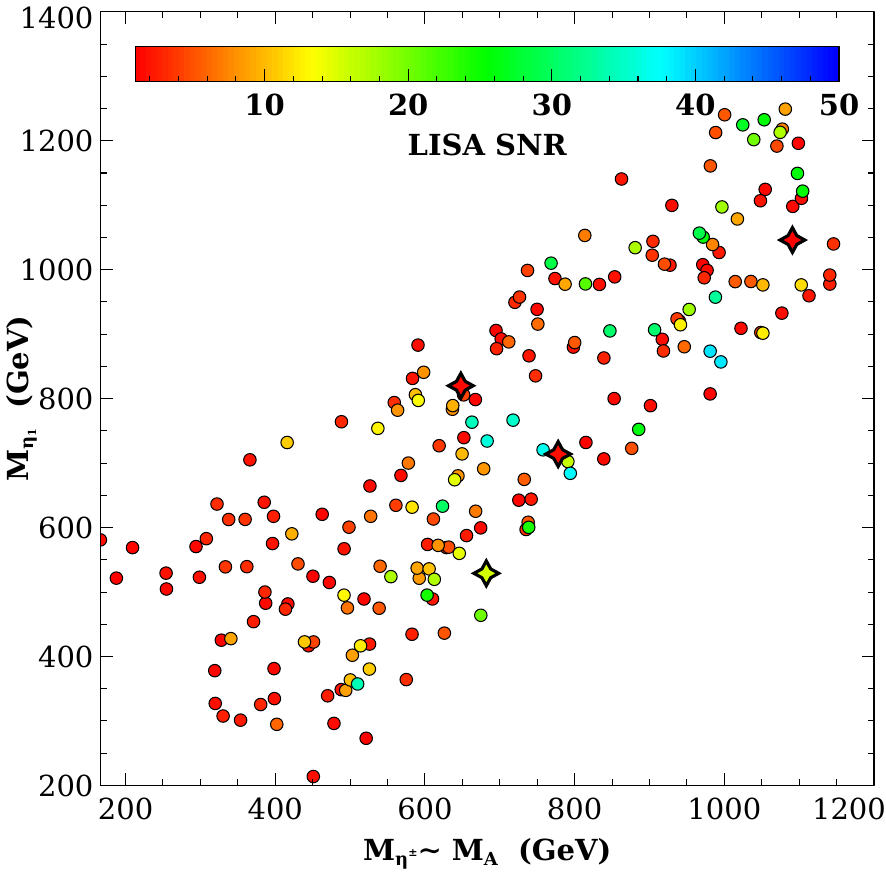}
    \includegraphics[width=0.48\linewidth]{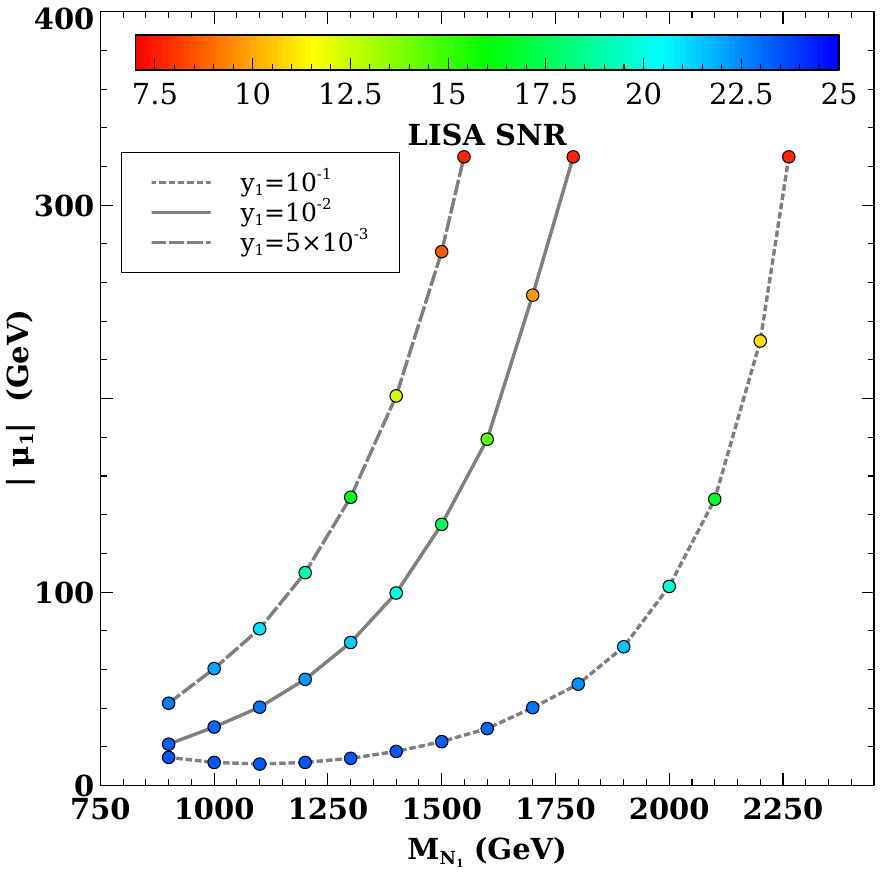}
    \caption{Left panel: Parameter space in terms of physical $Z_2$-odd scalar masses consistent with a strong first-order EWPT. The \ding{71}-shaped points correspond to the four benchmark points in table \ref{tab1}. Right panel: Parameter space in $M_{N_1}$-$\lvert \mu_1 \rvert$ plane with varying $y_1$ consistent with observed baryon asymmetry. For both the panels, other relevant parameters are kept fixed as BP1 in table \ref{tab1} and the colour code indicates the SNR for LISA.}
    \label{fig:scan}
\end{figure}

%\begin{figure}
%    \centering
%    \includegraphics[width=0.5\linewidth]{Plot1.pdf}
%    \caption{Baryon asymmetry satisfied $M_N$ vs $\mu_1$ plane with LISA signal to noise ratio (SNR) in color code: $M_\psi=10^4$ GeV, $y_1=10^{-3}$ $\mu_\chi=430$ GeV, $\mu_\eta=450$ GeV, $M_H=529$ GeV, $M_A =M_\eta=682$ GeV, $M_\chi=600$ GeV, $\lambda_5=0.1$.}
%    \label{fig:enter-label}
%\end{figure}

\section{Results and Discussion}
\label{sec4}
We have implemented the model in \texttt{CalcHEP}~\cite{Belyaev:2012qa} and \texttt{micrOMEGAs}~\cite{Alguero:2023zol} for the purpose of numerical calculations. Table \ref{tab1} lists a few benchmark parameters of the model consistent with the observed baryon asymmetry and DM relic. Fig. \ref{fig:lepto} shows the evolution of comoving abundances $Y_{N_1}, Y_{B-L}$ for different benchmark parameters connecting $N_1$ and SM to the dark sector. As Dirac Yukawa coupling of TeV scale $N_1$ to neutrinos are small due to constraints from light neutrino data, we utilize the freedom in choosing $y_1, \mu_1$ to enhance the CP asymmetry. The left and right panels of Fig. \ref{fig:lepto} show the variation of lepton asymmetry for variations in $y_1$ and $\mu_1$ respectively. The other relevant parameters are kept same as BP1 in table \ref{tab1}. While variation in $\mu_1$ can affect the masses of physical scalars $\eta_{1,2}$, the scalar portal couplings are varied to keep them fixed as the ones in BP1 when $\mu_1$ is being varied. As expected, due to the enhancement in CP asymmetry with larger $y_1, \mu_1$, the corresponding lepton asymmetry also rises. The larger Yukawa coupling $y_1$ compared to Dirac Yukawa coupling $y_N$ also brings $N_1$ into equilibrium very fast resulting in the sharp increase in $Y_{N_1}$ at high temperature. While CP asymmetry is independent of leptonic Dirac CP phase upon summing over flavours, we consider $z=0.5$ such that the orthogonal matrix $R$ does not provide any source of CP violation. We consider the $\mu_1$ term to be complex and hence the only source of CP phase going into the production of lepton asymmetry. Fig. \ref{fig:DM} shows the variation in DM abundance for different values of $\mu_1$ keeping other relevant parameters same as BP1 in table \ref{tab1}. Increase in $\mu_1$ leads to increase in DM ($\eta_1$ in this case) coupling to the SM Higgs enhancing its annihilation into a pair of Higgs. This results in smaller freeze-out relic of $\eta_1$ for larger $\mu_1$, as can be seen from Fig. \ref{fig:DM}.

We then check the possibility of a first-order EWPT for the chosen benchmark parameters in table \ref{tab1}, consistent with the observed baryon asymmetry and dark matter. The left panel of Fig. \ref{fig:gw} shows the profile of the finite-temperature effective potential at three different temperatures namely, $T=0, T_n, T_c$ for benchmark choice of parameters denoted by BP1 in table \ref{tab1}. Clearly, as we increase temperature from $T=0$ to $T=T_c$, the second degenerate minima appears confirming the first-order phase transition. The right panel plot of Fig. \ref{fig:gw} shows the GW spectra resulting from first-order EWPT for the benchmark points given in table \ref{tab1}. The experimental sensitivities of GW detectors BBO~\cite{Crowder:2005nr,Corbin:2005ny,Harry:2006fi}, DECIGO~\cite{Seto:2001qf,Kawamura:2006up,Yagi:2011wg}, ET~\cite{Punturo:2010zz, Hild:2010id, Sathyaprakash:2012jk, ET:2019dnz}, LISA~\cite{2017arXiv170200786A}, $\mu$ARES \cite{Sesana:2019vho} and THEIA~\cite{Garcia-Bellido:2021zgu} are shown as shaded regions of different colours.

\begin{figure}
    \centering
   \includegraphics[width=0.5\linewidth]{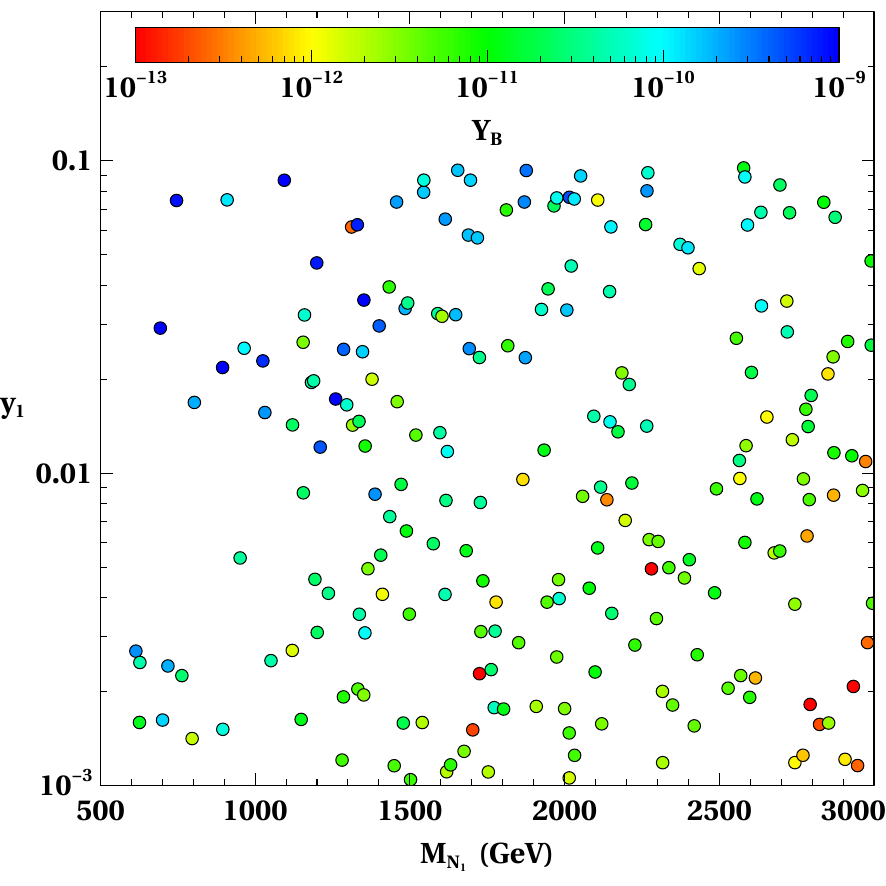}
    \caption{Parameter space in $y_1-M_{N_1}$ plane consistent with FOPT with colour code indicating the resulting baryon asymmetry. The other parameters are kept same as in the left panel of Fig. \ref{fig:scan}.}
    \label{fig:scan1}
\end{figure}

We also calculate the signal-to-noise ratio (SNR) for GW at LISA detector and show its variation via colour code in Fig. \ref{fig:scan}. The SNR is defined as~\cite{Schmitz:2020syl} 
\begin{equation}
\rho = \sqrt{\tau\,\int_{f_\text{min}}^{f_\text{max}}\,df\,\left[\frac{\Omega_\text{GW}(f)\,h^2}{\Omega_\text{expt}(f)\,h^2}\right]^2}\,, 
\end{equation}
with $\tau$ being the observation time for a particular detector, which we consider to be 5 yrs. The left panel of Fig. \ref{fig:scan} show the parameter space in terms of dark scalar masses $M_{\eta_1}, M_{\eta^\pm} \sim M_A$ consistent with a first-order EWPT while the colour code indicating the corresponding SNR for LISA. The \ding{71}-shaped points in the left panel of Fig. \ref{fig:scan} refer to the benchmark points in table \ref{tab1} which are consistent with the observed baryon asymmetry and dark matter abundance. The right panel plot of the same figure shows the parameter space in $\lvert \mu_1 \rvert-M_{N_1}$ plane with varying $y_1$ such that the criteria for the observed baryon asymmetry is satisfied. The colour code in this plot also indicates the SNR for LISA. While varying $\lvert \mu_1 \rvert, M_{N_1}, y_1$ in the right panel plot of Fig. \ref{fig:scan}, other relevant parameters are kept same as BP1 in table \ref{tab1}. As CP asymmetry is proportional to $y_1 \mu_1$, a smaller $y_1$ requires a larger $\mu_1$ and vice versa for a fixed scale of leptogenesis. This can be seen from the correlations shown in the right panel of Fig. \ref{fig:scan}. For the points satisfying FOPT criteria shown in the left panel of Fig. \ref{fig:scan}, we perform a random scan by varying $y_1 \in (10^{-3}, 0.1)$ and  $M_{N_1} \in (\mu_\eta+\mu_\chi, 3 \, {\rm TeV})$ to find out the resulting baryon asymmetry. The parameter space is shown in Fig. \ref{fig:scan1} with colour code indicating $Y_B (T_0)$. As noted earlier, a larger $y_1$ enhances the CP asymmetry and hence results in a larger value of baryon asymmetry. It should be noted that while fulfilling the requirements for the observed baryon asymmetry, dark matter and a first-order EWPT, we always fit the light parameters to light neutrino data by using the Casas-Ibarra parametrisation mentioned earlier. For example, the benchmark point BP1, the Dirac Yukawa coupling matrix given in Eq. \eqref{eq:yuk} turns out to be
\begin{equation}
\small
    \mathcal{Y}=\begin{pmatrix}
       -2.53\times10^{-7}-1.91\times10^{-6}i  & 8.59\times10^{-7}-7.50\times10^{-6}i  & 1.23\times10^{-7}-2.08\times10^{-6}i \\
2.72\times10^{-7}+1.27\times10^{-7}i  & -5.99\times10^{-7}-8.83\times10^{-8}i  & -9.48\times10^{-7}-1.35\times10^{-7}i  
    \end{pmatrix}. \nonumber 
\end{equation}

Due to the sub-TeV particle spectrum, the model can also have other interesting signatures. The new scalars in the model can can give rise to same-sign dilepton plus missing energy \cite{Gustafsson:2012aj, Datta:2016nfz}, dijet plus missing energy \cite{Poulose:2016lvz}, tri-lepton plus missing energy \cite{Miao:2010rg} or even mono jet signatures \cite{Belyaev:2016lok, Belyaev:2018ext} in colliders. Light scalar DM window can also be probed via precise measurements of the Higgs invisible branching ratio. Similar to the scotogenic model, we can also have interesting prospects of charged lepton flavour violating decays like $\mu \rightarrow e \gamma, \mu \rightarrow 3e$ due to light $N_1, \eta$ going inside the loop mediating such rare processes \cite{Toma:2013zsa}. The scalar DM in the model can also have sizeable DM-nucleon scattering rate mediated by Higgs or electroweak gauge bosons which can show up in direct detection experiments. Such DM can also annihilate into SM final states in local neighborhood opening up promising indirect detection prospects. The model also predicts vanishing lightest active neutrino mass keeping the effective neutrino mass much out of reach from ongoing tritium beta decay experiments like KATRIN \cite{KATRIN:2019yun}. Additionally, near future observation of neutrinoless double beta decay can also falsify our scenario, particularly for normal ordering of light neutrinos. The details of such associated phenomenology is beyond the scope of the present work and can be found elsewhere.

\section{Conclusion
\label{sec5}}
We have proposed a minimal framework for leptogenesis with associated gravitational wave signatures due to a strong first-order electroweak phase transition. While one heavy right-handed neutrino $N_1$ couples to lepton and Higgs doublet generating one of the light neutrino masses, another light neutrino mass is generated radiatively via a $Z_2$-odd sector, similar to the scoto-seesaw scenarios. While the simplest scoto-seesaw scenario can not provide non-zero CP asymmetry from right-handed neutrino decay at one-loop level, we extend it by a $Z_2$-odd singlet scalar generating a vertex correction to tree level decay width. The strength of the non-zero CP asymmetry depends upon the coupling of $N_1$ and the SM to the $Z_2$-odd sector. The $Z_2$-odd scalar mass and their coupling with the SM Higgs also drives the electroweak phase transition to a first-order phase transition with promising gravitational wave signatures. Due to the freedom in choosing dark sector couplings with $N_1$ and SM, we show the viability of TeV scale leptogenesis without any resonantly enhanced self-energy correction while at the same time being consistent with light neutrino data. While the lightest active neutrino mass is zero in this minimal setup, the hierarchical solar and atmospheric mass scales naturally arise from radiative and tree-level contributions respectively to neutrino mass. Dark matter is preferentially the lightest $Z_2$-odd scalar as $Z_2$-odd heavy fermion is assumed to be heavier than $N_1$ to enhance the production of lepton asymmetry. While the gravitational wave signal can be probed at several future experiments like LISA, the new particle degrees of freedom sub-TeV scale can lead to observable signatures related to collider, lepton flavour violation, dark matter detection etc. keeping the framework verifiable in near future.\\

\noindent
{\bf Note Added:} While this work was being completed, a related work \cite{Liu:2025xvm} appeared where the original Higgs portal leptogenesis model \cite{LeDall:2014too, Alanne:2018brf} with type-I seesaw and no dark matter has been studied in the context of a first-order phase transition and gravitational waves.

\section*{Acknowledgement}
The work of D.B. is supported by the Science and Engineering Research Board (SERB), Government of India grants MTR/2022/000575 and CRG/2022/000603. D.B. also acknowledges the support from the Fulbright-Nehru Academic and Professional Excellence Award 2024-25. 

\appendix

\section{Field dependent and thermal masses}
\label{appen1}
The squared field dependent masses with corresponding dof, relevant for the FOPT calculations, are

\begin{align}
     m_{\eta^\pm}^2 (\phi)&=\mu_\eta^2 +\frac{\lambda_3}{2}\phi^2 \,\,\, (n_{\eta^\pm} =2, C_{\eta^\pm} =\frac{3}{2}), \,\,\, m_A^2 (\phi)=\mu_\eta^2 +\frac{\lambda_3+\lambda_4 -2\lambda_5}{2}\phi^2 \, (n_A=1, C_{A} =\frac{3}{2}),
     \nonumber \\
    m_W^2(\phi)& =\frac{g_2^2}{4}\phi^2 \, (n_W=6, C_{W} =\frac{5}{6}), \,\,\, m_Z^2(\phi)=\frac{g_1^2+g_2^2}{4}\phi^2 \,\,\, (n_Z=3, C_{Z} =\frac{5}{6}), \nonumber \\
    m_t^2(\phi) & =\frac{y_t^2}{2}\phi^2 \,\,\, (n_t=12, C_{t} =\frac{3}{2}), \,\,\, m_b^2(\phi)=\frac{y_b^2}{2}\phi^2 \,\,\, (n_b=12, C_{b} =\frac{3}{2}), \nonumber\\
    m_{\eta_1}^2 (\phi) & = \frac{1}{2}(\mu_\chi^2+\mu_\eta^2+(\lambda_8+\lambda_{\rm H}/2)\phi^2-\sqrt{\{\mu_\chi^2-\mu_\eta^2+(\lambda_8-\lambda_{\rm H}/2)\phi^2 \}^2+\lvert \mu_1 \rvert^2\phi^2 }) \,\,\, (n_{\eta_1}=1, C_{\eta_1} =\frac{3}{2}),\nonumber \\
    m_{\eta_2}^2 (\phi) &= \frac{1}{2}(\mu_\chi^2+\mu_\eta^2+(\lambda_8+\lambda_{\rm H}/2)\phi^2+\sqrt{\{\mu_\chi^2-\mu_\eta^2+(\lambda_8-\lambda_{\rm H}/2)\phi^2 \}^2+\lvert \mu_1 \rvert^2\phi^2 }) \,\,\, (n_{\eta_2}=1, C_{\eta_2} =\frac{3}{2}). \nonumber
\end{align}

The thermal masses for inert doublet components and scalar singlet are $m^2_i(\phi,T)=m^2_i(\phi) + \Pi_S(T),\, m^2_{\eta_2}(\phi,T)=m^2_{\eta_2}(\phi) + \Pi_\chi(T)$ respectively. On the other hand, the thermal masses for electroweak vector bosons are 
\begin{align}
m_{W_L}^2(\phi,T) = m_W^2(\phi) +\Pi_W(T), \nonumber \\
m_{Z_L}^2(\phi,T)=\frac{1}{2}( m_Z^2(\phi) +\Pi_W(T)+ \Pi_Y(T)+\Delta(\phi,T)), \nonumber \\
 m_{\gamma_L}^2(\phi,T)=\frac{1}{2}( m_Z^2(\phi) +\Pi_W(T)+ \Pi_Y(T)-\Delta(\phi,T))
\end{align}
where
\begin{align}
\Pi_S(T) & = \bigg (\frac{1}{8}g_2^2+\frac{1}{16}(g_1^2+g_2^2)+\frac{1}{2}\lambda_2+\frac{1}{12}\lambda_3+\frac{1}{24}\lambda_A +\frac{1}{24}\lambda_H +\frac{1}{12}\lambda_9 \bigg )T^2 \nonumber \\
\Pi_\chi(T)& =\bigg ( \frac{1}{4}\lambda_7+\frac{1}{3}\lambda_8+\frac{1}{3}\lambda_9 \bigg)T^2,
 \Pi_W(T)=2g_2^2T^2, \,\, \Pi_Y(T)=2g_1^2T^2, \nonumber \\
\lambda_H & =\lambda_3+\lambda_4 +2\lambda_5, \,\, \lambda_A=\lambda_3+\lambda_4 -2\lambda_5.
\end{align}
The singlet fermions $N_1, \psi$ do not receive much thermal corrections and hence their masses are same as their bare masses.

\section{Gravitational waves from first-order phase transition}
\label{appen2}
The GW spectrum for bubble collision is given by \cite{Caprini:2015zlo}
\begin{equation}
    \Omega_\phi h^2 = 1.67 \times 10^{-5} \left ( \frac{100}{g_*} \right)^{1/3} \left(\frac{{ H_*}}{\beta}\right)^2 \left(\frac{\kappa_\phi \alpha_*}{1+\alpha_*}\right)^2 \frac{0.11 v^3_w}{0.42+v^2_w} \frac{3.8(f/f_{\rm peak}^{\rm PT, \phi})^{2.8}}{1+2.8 (f/f_{\rm peak}^{\rm PT, \phi})^{3.8}}\,,
\end{equation}
where the peak frequency \cite{Caprini:2015zlo} is 
\begin{equation}
    f_{\rm peak}^{\rm PT, \phi} = 1.65 \times 10^{-5} {\rm Hz} \left ( \frac{g_*}{100} \right)^{1/6} \left ( \frac{T_n}{100 \; {\rm GeV}} \right ) \frac{0.62}{1.8-0.1v_w+v^2_w} \left(\frac{\beta}{{ H_*}}\right).
\end{equation}
The Hubble expansion parameter at nucleation temperature is denoted as $H_*=H(T_n)$. The efficiency factor $\kappa_\phi$ for bubble collision can be expressed as \cite{Kamionkowski:1993fg}
\begin{align}
    \kappa_\phi=\frac{1}{1+0.715 \alpha_*}\left(0.715\alpha_* +\frac{4}{27}\sqrt{3\alpha_*/2}\right).
\end{align}
The GW spectrum produced from the sound wave in the plasma can be written as \cite{Caprini:2015zlo,Caprini:2019egz,Guo:2020grp}
\begin{equation}
    \Omega_{\rm sw} h^2 = 2.65 \times 10^{-6} \left ( \frac{100}{g_*} \right)^{1/3} \left(\frac{{ H_*}}{\beta}\right) \left(\frac{\kappa_{\rm sw} \alpha_*}{1+\alpha_*}\right)^2 v_w (f/f_{\rm peak}^{\rm PT, sw})^{3} \left(\frac{7}{4+3 (f/f_{\rm peak}^{\rm PT, sw})^{2}} \right)^{7/2} \Upsilon
\end{equation}
and the corresponding peak frequency is given by \cite{Caprini:2015zlo}
\begin{equation}
    f_{\rm peak}^{\rm PT, sw} = 1.65 \times 10^{-5} {\rm Hz} \left ( \frac{g_*}{100} \right)^{1/6} \left ( \frac{T_n}{100 \; {\rm GeV}} \right )  \left(\frac{\beta}{{ H_*}}\right) \frac{2}{\sqrt{3}}.
\end{equation}
The efficiency factor for sound wave can be expressed as \cite{Espinosa:2010hh}
\begin{align}
    \kappa_{\rm sw}=\frac{\sqrt{\alpha_*}}{0.135+ \sqrt{0.98+\alpha_*}}.
\end{align}
The suppression factor $\Upsilon=1-\frac{1}{\sqrt{1+2\tau_{sw}H_*}}$ depends on the lifetime of sound wave $\tau_{\rm sw}$\cite{Guo:2020grp} given by $\tau_{\rm sw}\sim R_*/\bar{U}_f$ where mean bubble separation is $R_*=(8\pi)^{1/3}v_w \beta^{-1}$ and rms fluid velocity is $\bar{U}_f=\sqrt{3\kappa_{sw}\alpha_*/4}$.
Finally, the GW spectrum generated by the turbulence in the plasma is given by \cite{Caprini:2015zlo}
\begin{equation}
    \Omega_{\rm turb} h^2 = 3.35 \times 10^{-4} \left ( \frac{100}{g_*} \right)^{1/3} \left(\frac{{ H_*}}{\beta}\right) \left(\frac{\kappa_{\rm turb} \alpha_*}{1+\alpha_*}\right)^{3/2} v_w \frac{(f/f_{\rm peak}^{\rm PT, turb})^{3}}{(1+ f/f_{\rm peak}^{\rm PT, turb})^{11/3} (1+8\pi f/h_*)}
\end{equation}
with the peak frequency as \cite{Caprini:2015zlo}
\begin{equation}
    f_{\rm peak}^{\rm PT, turb} = 1.65 \times 10^{-5} {\rm Hz} \left ( \frac{g_*}{100} \right)^{1/6} \left ( \frac{T_n}{100 \; {\rm GeV}} \right ) \frac{3.5}{2} \left(\frac{\beta}{{H_*}}\right).
\end{equation}
The efficiency factor for turbulence is $\kappa_{\rm turb}\simeq 0.1\kappa_{\rm sw}$ and the inverse Hubble time at the epoch of GW emission, redshifted to today is
\begin{equation}
   h_*= 1.65\times 10^{-5} \frac{T_n}{100 \hspace{0.1 cm} \rm GeV} \left(\frac{g_*}{100}\right)^{1/6}.
\end{equation}
 It is clear from the above expressions that the contribution from sound waves turns out to be the most dominant one and the peak of the total GW spectrum corresponds to the peak frequency of sound waves contribution.

%\bibliographystyle{JHEP}
%\bibliography{refa, refb, refc}

\begin{thebibliography}{100}

\bibitem{Zyla:2020zbs}
{\bf Particle Data Group} Collaboration, P.~A. Zyla {\em et~al.}, {\it {Review of Particle Physics}},  {\em PTEP} {\bf 2020} (2020), no.~8 083C01.

\bibitem{Aghanim:2018eyx}
{\bf Planck} Collaboration, N.~Aghanim {\em et~al.}, {\it {Planck 2018 results. VI. Cosmological parameters}},  \href{http://arxiv.org/abs/1807.06209}{{\tt arXiv:1807.06209}}.

\bibitem{Sakharov:1967dj}
A.~D. Sakharov, {\it {Violation of CP Invariance, C asymmetry, and baryon asymmetry of the universe}},  {\em Pisma Zh. Eksp. Teor. Fiz.} {\bf 5} (1967) 32--35. [Usp. Fiz. Nauk161,no.5,61(1991)].

\bibitem{Kolb:1990vq}
E.~W. Kolb and M.~S. Turner, {\it {The Early Universe}},  {\em Front. Phys.} {\bf 69} (1990) 1--547.

\bibitem{Jungman:1995df}
G.~Jungman, M.~Kamionkowski, and K.~Griest, {\it {Supersymmetric dark matter}},  {\em Phys. Rept.} {\bf 267} (1996) 195--373, [\href{http://arxiv.org/abs/hep-ph/9506380}{{\tt hep-ph/9506380}}].

\bibitem{Weinberg:1979bt}
S.~Weinberg, {\it {Cosmological Production of Baryons}},  {\em Phys. Rev. Lett.} {\bf 42} (1979) 850--853.

\bibitem{Kolb:1979qa}
E.~W. Kolb and S.~Wolfram, {\it {Baryon Number Generation in the Early Universe}},  {\em Nucl. Phys.} {\bf B172} (1980) 224. [Erratum: Nucl. Phys.B195,542(1982)].

\bibitem{Fukugita:1986hr}
M.~Fukugita and T.~Yanagida, {\it {Baryogenesis Without Grand Unification}},  {\em Phys. Lett.} {\bf B174} (1986) 45--47.

\bibitem{Kuzmin:1985mm}
V.~A. Kuzmin, V.~A. Rubakov, and M.~E. Shaposhnikov, {\it {On the Anomalous Electroweak Baryon Number Nonconservation in the Early Universe}},  {\em Phys. Lett.} {\bf 155B} (1985) 36.

\bibitem{Minkowski:1977sc}
P.~Minkowski, {\it {$\mu \to e\gamma$ at a Rate of One Out of $10^{9}$ Muon Decays?}},  {\em Phys. Lett. B} {\bf 67} (1977) 421--428.

\bibitem{GellMann:1980vs}
M.~Gell-Mann, P.~Ramond, and R.~Slansky, {\it {Complex Spinors and Unified Theories}},  {\em Conf. Proc.} {\bf C790927} (1979) 315--321, [\href{http://arxiv.org/abs/1306.4669}{{\tt arXiv:1306.4669}}].

\bibitem{Mohapatra:1979ia}
R.~N. Mohapatra and G.~Senjanovic, {\it {Neutrino Mass and Spontaneous Parity Violation}},  {\em Phys. Rev. Lett.} {\bf 44} (1980) 912.

\bibitem{Schechter:1980gr}
J.~Schechter and J.~W.~F. Valle, {\it {Neutrino Masses in SU(2) x U(1) Theories}},  {\em Phys. Rev.} {\bf D22} (1980) 2227.

\bibitem{Schechter:1981cv}
J.~Schechter and J.~W.~F. Valle, {\it {Neutrino Decay and Spontaneous Violation of Lepton Number}},  {\em Phys. Rev. D} {\bf 25} (1982) 774.

\bibitem{LZ:2022lsv}
{\bf LZ} Collaboration, J.~Aalbers {\em et~al.}, {\it {First Dark Matter Search Results from the LUX-ZEPLIN (LZ) Experiment}},  {\em Phys. Rev. Lett.} {\bf 131} (2023), no.~4 041002, [\href{http://arxiv.org/abs/2207.03764}{{\tt arXiv:2207.03764}}].

\bibitem{Hall:2019ank}
E.~Hall, T.~Konstandin, R.~McGehee, H.~Murayama, and G.~Servant, {\it {Baryogenesis From a Dark First-Order Phase Transition}},  {\em JHEP} {\bf 04} (2020) 042, [\href{http://arxiv.org/abs/1910.08068}{{\tt arXiv:1910.08068}}].

\bibitem{Dror:2019syi}
J.~A. Dror, T.~Hiramatsu, K.~Kohri, H.~Murayama, and G.~White, {\it {Testing the Seesaw Mechanism and Leptogenesis with Gravitational Waves}},  {\em Phys. Rev. Lett.} {\bf 124} (2020), no.~4 041804, [\href{http://arxiv.org/abs/1908.03227}{{\tt arXiv:1908.03227}}].

\bibitem{Blasi:2020wpy}
S.~Blasi, V.~Brdar, and K.~Schmitz, {\it {Fingerprint of low-scale leptogenesis in the primordial gravitational-wave spectrum}},  {\em Phys. Rev. Res.} {\bf 2} (2020), no.~4 043321, [\href{http://arxiv.org/abs/2004.02889}{{\tt arXiv:2004.02889}}].

\bibitem{Fornal:2020esl}
B.~Fornal and B.~Shams Es~Haghi, {\it {Baryon and Lepton Number Violation from Gravitational Waves}},  {\em Phys. Rev. D} {\bf 102} (2020), no.~11 115037, [\href{http://arxiv.org/abs/2008.05111}{{\tt arXiv:2008.05111}}].

\bibitem{Samanta:2020cdk}
R.~Samanta and S.~Datta, {\it {Gravitational wave complementarity and impact of NANOGrav data on gravitational leptogenesis: cosmic strings}},  \href{http://arxiv.org/abs/2009.13452}{{\tt arXiv:2009.13452}}.

\bibitem{Barman:2022yos}
B.~Barman, D.~Borah, A.~Dasgupta, and A.~Ghoshal, {\it {Probing high scale Dirac leptogenesis via gravitational waves from domain walls}},  {\em Phys. Rev. D} {\bf 106} (2022), no.~1 015007, [\href{http://arxiv.org/abs/2205.03422}{{\tt arXiv:2205.03422}}].

\bibitem{Baldes:2021vyz}
I.~Baldes, S.~Blasi, A.~Mariotti, A.~Sevrin, and K.~Turbang, {\it {Baryogenesis via relativistic bubble expansion}},  {\em Phys. Rev. D} {\bf 104} (2021), no.~11 115029, [\href{http://arxiv.org/abs/2106.15602}{{\tt arXiv:2106.15602}}].

\bibitem{Azatov:2021irb}
A.~Azatov, M.~Vanvlasselaer, and W.~Yin, {\it {Baryogenesis via relativistic bubble walls}},  {\em JHEP} {\bf 10} (2021) 043, [\href{http://arxiv.org/abs/2106.14913}{{\tt arXiv:2106.14913}}].

\bibitem{Huang:2022vkf}
P.~Huang and K.-P. Xie, {\it {Leptogenesis triggered by a first-order phase transition}},  {\em JHEP} {\bf 09} (2022) 052, [\href{http://arxiv.org/abs/2206.04691}{{\tt arXiv:2206.04691}}].

\bibitem{Dasgupta:2022isg}
A.~Dasgupta, P.~S.~B. Dev, A.~Ghoshal, and A.~Mazumdar, {\it {Gravitational wave pathway to testable leptogenesis}},  {\em Phys. Rev. D} {\bf 106} (2022), no.~7 075027, [\href{http://arxiv.org/abs/2206.07032}{{\tt arXiv:2206.07032}}].

\bibitem{Barman:2022pdo}
B.~Barman, D.~Borah, S.~Jyoti~Das, and R.~Roshan, {\it {Gravitational wave signatures of a PBH-generated baryon-dark matter coincidence}},  {\em Phys. Rev. D} {\bf 107} (2023), no.~9 095002, [\href{http://arxiv.org/abs/2212.00052}{{\tt arXiv:2212.00052}}].

\bibitem{Datta:2022tab}
S.~Datta and R.~Samanta, {\it {Gravitational waves-tomography of Low-Scale-Leptogenesis}},  {\em JHEP} {\bf 11} (2022) 159, [\href{http://arxiv.org/abs/2208.09949}{{\tt arXiv:2208.09949}}].

\bibitem{Borah:2022cdx}
D.~Borah, A.~Dasgupta, and I.~Saha, {\it {Leptogenesis and dark matter through relativistic bubble walls with observable gravitational waves}},  {\em JHEP} {\bf 11} (2022) 136, [\href{http://arxiv.org/abs/2207.14226}{{\tt arXiv:2207.14226}}].

\bibitem{Borah:2023saq}
D.~Borah, A.~Dasgupta, and I.~Saha, {\it {LIGO-VIRGO constraints on dark matter and leptogenesis triggered by a first order phase transition at high scale}},  \href{http://arxiv.org/abs/2304.08888}{{\tt arXiv:2304.08888}}.

\bibitem{Borah:2023god}
D.~Borah, A.~Dasgupta, M.~Knauss, and I.~Saha, {\it {Baryon asymmetry from dark matter decay in the vicinity of a phase transition}},  {\em Phys. Rev. D} {\bf 108} (2023), no.~9 L091701, [\href{http://arxiv.org/abs/2306.05459}{{\tt arXiv:2306.05459}}].

\bibitem{Barman:2023fad}
B.~Barman, D.~Borah, S.~Jyoti~Das, and I.~Saha, {\it {Scale of Dirac leptogenesis and left-right symmetry in the light of recent PTA results}},  {\em JCAP} {\bf 10} (2023) 053, [\href{http://arxiv.org/abs/2307.00656}{{\tt arXiv:2307.00656}}].

\bibitem{Borah:2024qyo}
D.~Borah, N.~Das, S.~Jyoti~Das, and R.~Samanta, {\it {Cogenesis of baryon and dark matter with PBHs and the QCD axion}},  {\em Phys. Rev. D} {\bf 110} (2024), no.~11 115013, [\href{http://arxiv.org/abs/2403.02401}{{\tt arXiv:2403.02401}}].

\bibitem{Borah:2024bcr}
D.~Borah and N.~Das, {\it {Successful cogenesis of baryon and dark matter from memory-burdened PBH}},  {\em JCAP} {\bf 02} (2025) 031, [\href{http://arxiv.org/abs/2410.16403}{{\tt arXiv:2410.16403}}].

\bibitem{Barman:2024ujh}
B.~Barman, A.~Basu, D.~Borah, A.~Chakraborty, and R.~Roshan, {\it {Testing leptogenesis and dark matter production during reheating with primordial gravitational waves}},  {\em Phys. Rev. D} {\bf 111} (2025), no.~5 055016, [\href{http://arxiv.org/abs/2410.19048}{{\tt arXiv:2410.19048}}].

\bibitem{Hall:2019rld}
E.~Hall, T.~Konstandin, R.~McGehee, and H.~Murayama, {\it {Asymmetric Matters from a Dark First-Order Phase Transition}},  \href{http://arxiv.org/abs/1911.12342}{{\tt arXiv:1911.12342}}.

\bibitem{Yuan:2021ebu}
C.~Yuan, R.~Brito, and V.~Cardoso, {\it {Probing ultralight dark matter with future ground-based gravitational-wave detectors}},  {\em Phys. Rev. D} {\bf 104} (2021), no.~4 044011, [\href{http://arxiv.org/abs/2106.00021}{{\tt arXiv:2106.00021}}].

\bibitem{Tsukada:2020lgt}
L.~Tsukada, R.~Brito, W.~E. East, and N.~Siemonsen, {\it {Modeling and searching for a stochastic gravitational-wave background from ultralight vector bosons}},  {\em Phys. Rev. D} {\bf 103} (2021), no.~8 083005, [\href{http://arxiv.org/abs/2011.06995}{{\tt arXiv:2011.06995}}].

\bibitem{Chatrchyan:2020pzh}
A.~Chatrchyan and J.~Jaeckel, {\it {Gravitational waves from the fragmentation of axion-like particle dark matter}},  {\em JCAP} {\bf 02} (2021) 003, [\href{http://arxiv.org/abs/2004.07844}{{\tt arXiv:2004.07844}}].

\bibitem{Bian:2021vmi}
L.~Bian, X.~Liu, and K.-P. Xie, {\it {Probing superheavy dark matter with gravitational waves}},  {\em JHEP} {\bf 11} (2021) 175, [\href{http://arxiv.org/abs/2107.13112}{{\tt arXiv:2107.13112}}].

\bibitem{Samanta:2021mdm}
R.~Samanta and F.~R. Urban, {\it {Testing Super Heavy Dark Matter from Primordial Black Holes with Gravitational Waves}},  \href{http://arxiv.org/abs/2112.04836}{{\tt arXiv:2112.04836}}.

\bibitem{Borah:2022byb}
D.~Borah, S.~J. Das, A.~K. Saha, and R.~Samanta, {\it {Probing Miracle-less WIMP Dark Matter via Gravitational Waves Spectral Shapes}},  \href{http://arxiv.org/abs/2202.10474}{{\tt arXiv:2202.10474}}.

\bibitem{Azatov:2021ifm}
A.~Azatov, M.~Vanvlasselaer, and W.~Yin, {\it {Dark Matter production from relativistic bubble walls}},  {\em JHEP} {\bf 03} (2021) 288, [\href{http://arxiv.org/abs/2101.05721}{{\tt arXiv:2101.05721}}].

\bibitem{Azatov:2022tii}
A.~Azatov, G.~Barni, S.~Chakraborty, M.~Vanvlasselaer, and W.~Yin, {\it {Ultra-relativistic bubbles from the simplest Higgs portal and their cosmological consequences}},  {\em JHEP} {\bf 10} (2022) 017, [\href{http://arxiv.org/abs/2207.02230}{{\tt arXiv:2207.02230}}].

\bibitem{Baldes:2022oev}
I.~Baldes, Y.~Gouttenoire, and F.~Sala, {\it {Hot and heavy dark matter from a weak scale phase transition}},  {\em SciPost Phys.} {\bf 14} (2023), no.~3 033, [\href{http://arxiv.org/abs/2207.05096}{{\tt arXiv:2207.05096}}].

\bibitem{Borah:2022iym}
D.~Borah, S.~Jyoti~Das, R.~Samanta, and F.~R. Urban, {\it {PBH-infused seesaw origin of matter and unique gravitational waves}},  {\em JHEP} {\bf 03} (2023) 127, [\href{http://arxiv.org/abs/2211.15726}{{\tt arXiv:2211.15726}}].

\bibitem{Borah:2022vsu}
D.~Borah, S.~Jyoti~Das, and R.~Roshan, {\it {Probing high scale seesaw and PBH generated dark matter via gravitational waves with multiple tilts}},  {\em Nucl. Phys. B} {\bf 1002} (2024) 116528, [\href{http://arxiv.org/abs/2208.04965}{{\tt arXiv:2208.04965}}].

\bibitem{Shibuya:2022xkj}
H.~Shibuya and T.~Toma, {\it {Impact of first-order phase transitions on dark matter production in the scotogenic model}},  {\em JHEP} {\bf 11} (2022) 064, [\href{http://arxiv.org/abs/2207.14662}{{\tt arXiv:2207.14662}}].

\bibitem{Borah:2023sbc}
D.~Borah, S.~Jyoti~Das, and R.~Samanta, {\it {Imprint of inflationary gravitational waves and WIMP dark matter in pulsar timing array data}},  {\em JCAP} {\bf 03} (2024) 031, [\href{http://arxiv.org/abs/2307.00537}{{\tt arXiv:2307.00537}}].

\bibitem{Borah:2024lml}
D.~Borah, S.~Jyoti~Das, and I.~Saha, {\it {Dark matter from phase transition generated PBH evaporation with gravitational waves signatures}},  {\em Phys. Rev. D} {\bf 110} (2024), no.~3 035014, [\href{http://arxiv.org/abs/2401.12282}{{\tt arXiv:2401.12282}}].

\bibitem{Adhikary:2024btd}
A.~Adhikary, D.~Borah, S.~Mahapatra, I.~Saha, N.~Sahu, and V.~S. Thounaojam, {\it {New realisation of light thermal dark matter with enhanced detection prospects}},  {\em JCAP} {\bf 12} (2024) 043, [\href{http://arxiv.org/abs/2405.17564}{{\tt arXiv:2405.17564}}].

\bibitem{Borah:2024kfn}
D.~Borah, N.~Das, and R.~Roshan, {\it {Observable gravitational waves and \ensuremath{\Delta}Neff with global lepton number symmetry and dark matter}},  {\em Phys. Rev. D} {\bf 110} (2024), no.~7 075042, [\href{http://arxiv.org/abs/2406.04404}{{\tt arXiv:2406.04404}}].

\bibitem{Borboruah:2024lli}
Z.~A. Borboruah, D.~Borah, L.~Malhotra, and U.~Patel, {\it {Minimal Dirac seesaw dark matter}},  \href{http://arxiv.org/abs/2412.12267}{{\tt arXiv:2412.12267}}.

\bibitem{Borah:2025wzl}
D.~Borah and I.~Saha, {\it {Filtered cogenesis of PBH dark matter and baryons}},  \href{http://arxiv.org/abs/2502.12248}{{\tt arXiv:2502.12248}}.

\bibitem{LeDall:2014too}
M.~Le~Dall and A.~Ritz, {\it {Leptogenesis and the Higgs Portal}},  {\em Phys. Rev. D} {\bf 90} (2014), no.~9 096002, [\href{http://arxiv.org/abs/1408.2498}{{\tt arXiv:1408.2498}}].

\bibitem{Alanne:2018brf}
T.~Alanne, T.~Hugle, M.~Platscher, and K.~Schmitz, {\it {Low-scale leptogenesis assisted by a real scalar singlet}},  {\em JCAP} {\bf 03} (2019) 037, [\href{http://arxiv.org/abs/1812.04421}{{\tt arXiv:1812.04421}}].

\bibitem{Pilaftsis:2003gt}
A.~Pilaftsis and T.~E.~J. Underwood, {\it {Resonant leptogenesis}},  {\em Nucl. Phys.} {\bf B692} (2004) 303--345, [\href{http://arxiv.org/abs/hep-ph/0309342}{{\tt hep-ph/0309342}}].

\bibitem{Rojas:2018wym}
N.~Rojas, R.~Srivastava, and J.~W.~F. Valle, {\it {Simplest Scoto-Seesaw Mechanism}},  {\em Phys. Lett. B} {\bf 789} (2019) 132--136, [\href{http://arxiv.org/abs/1807.11447}{{\tt arXiv:1807.11447}}].

\bibitem{Tao:1996vb}
Z.-j. Tao, {\it {Radiative seesaw mechanism at weak scale}},  {\em Phys. Rev. D} {\bf 54} (1996) 5693--5697, [\href{http://arxiv.org/abs/hep-ph/9603309}{{\tt hep-ph/9603309}}].

\bibitem{Ma:2006km}
E.~Ma, {\it {Verifiable radiative seesaw mechanism of neutrino mass and dark matter}},  {\em Phys. Rev. D} {\bf 73} (2006) 077301, [\href{http://arxiv.org/abs/hep-ph/0601225}{{\tt hep-ph/0601225}}].

\bibitem{Beniwal:2020hjc}
A.~Beniwal, J.~Herrero-Garc\'\i{}a, N.~Leerdam, M.~White, and A.~G. Williams, {\it {The ScotoSinglet Model: a scalar singlet extension of the Scotogenic Model}},  {\em JHEP} {\bf 21} (2020) 136, [\href{http://arxiv.org/abs/2010.05937}{{\tt arXiv:2010.05937}}].

\bibitem{Casas:2001sr}
J.~A. Casas and A.~Ibarra, {\it {Oscillating neutrinos and $\mu \to e, \gamma$}},  {\em Nucl. Phys. B} {\bf 618} (2001) 171--204, [\href{http://arxiv.org/abs/hep-ph/0103065}{{\tt hep-ph/0103065}}].

\bibitem{Toma:2013zsa}
T.~Toma and A.~Vicente, {\it {Lepton Flavor Violation in the Scotogenic Model}},  {\em JHEP} {\bf 01} (2014) 160, [\href{http://arxiv.org/abs/1312.2840}{{\tt arXiv:1312.2840}}].

\bibitem{Leite:2023gzl}
J.~Leite, S.~Sadhukhan, and J.~W.~F. Valle, {\it {Dynamical scoto-seesaw mechanism with gauged B-L symmetry}},  {\em Phys. Rev. D} {\bf 109} (2024), no.~3 035023, [\href{http://arxiv.org/abs/2307.04840}{{\tt arXiv:2307.04840}}].

\bibitem{Ibarra:2003up}
A.~Ibarra and G.~G. Ross, {\it {Neutrino phenomenology: The Case of two right-handed neutrinos}},  {\em Phys. Lett. B} {\bf 591} (2004) 285--296, [\href{http://arxiv.org/abs/hep-ph/0312138}{{\tt hep-ph/0312138}}].

\bibitem{Bhattacharya:2024ohh}
S.~Bhattacharya, D.~Mahanta, N.~Mondal, and D.~Pradhan, {\it {Two-component Dark Matter and low scale Thermal Leptogenesis}},  \href{http://arxiv.org/abs/2412.21202}{{\tt arXiv:2412.21202}}.

\bibitem{Gondolo:1990dk}
P.~Gondolo and G.~Gelmini, {\it {Cosmic abundances of stable particles: Improved analysis}},  {\em Nucl. Phys.} {\bf B360} (1991) 145--179.

\bibitem{Buchmuller:2004nz}
W.~Buchmuller, P.~Di~Bari, and M.~Plumacher, {\it {Leptogenesis for pedestrians}},  {\em Annals Phys.} {\bf 315} (2005) 305--351, [\href{http://arxiv.org/abs/hep-ph/0401240}{{\tt hep-ph/0401240}}].

\bibitem{Harvey:1990qw}
J.~A. Harvey and M.~S. Turner, {\it {Cosmological baryon and lepton number in the presence of electroweak fermion number violation}},  {\em Phys. Rev. D} {\bf 42} (1990) 3344--3349.

\bibitem{Coleman:1973jx}
S.~R. Coleman and E.~J. Weinberg, {\it {Radiative Corrections as the Origin of Spontaneous Symmetry Breaking}},  {\em Phys. Rev. D} {\bf 7} (1973) 1888--1910.

\bibitem{Dolan:1973qd}
L.~Dolan and R.~Jackiw, {\it {Symmetry Behavior at Finite Temperature}},  {\em Phys. Rev. D} {\bf 9} (1974) 3320--3341.

\bibitem{Quiros:1999jp}
M.~Quiros, {\it {Finite temperature field theory and phase transitions}},  in {\em {ICTP Summer School in High-Energy Physics and Cosmology}}, pp.~187--259, 1, 1999.
\newblock \href{http://arxiv.org/abs/hep-ph/9901312}{{\tt hep-ph/9901312}}.

\bibitem{Mazumdar:2018dfl}
A.~Mazumdar and G.~White, {\it {Review of cosmic phase transitions: their significance and experimental signatures}},  {\em Rept. Prog. Phys.} {\bf 82} (2019), no.~7 076901, [\href{http://arxiv.org/abs/1811.01948}{{\tt arXiv:1811.01948}}].

\bibitem{Hindmarsh:2020hop}
M.~B. Hindmarsh, M.~L\"uben, J.~Lumma, and M.~Pauly, {\it {Phase transitions in the early universe}},  {\em SciPost Phys. Lect. Notes} {\bf 24} (2021) 1, [\href{http://arxiv.org/abs/2008.09136}{{\tt arXiv:2008.09136}}].

\bibitem{Athron:2023xlk}
P.~Athron, C.~Bal\'azs, A.~Fowlie, L.~Morris, and L.~Wu, {\it {Cosmological phase transitions: from perturbative particle physics to gravitational waves}},  \href{http://arxiv.org/abs/2305.02357}{{\tt arXiv:2305.02357}}.

\bibitem{Fendley:1987ef}
P.~Fendley, {\it {The Effective Potential and the Coupling Constant at High Temperature}},  {\em Phys. Lett. B} {\bf 196} (1987) 175--180.

\bibitem{Parwani:1991gq}
R.~R. Parwani, {\it {Resummation in a hot scalar field theory}},  {\em Phys. Rev. D} {\bf 45} (1992) 4695, [\href{http://arxiv.org/abs/hep-ph/9204216}{{\tt hep-ph/9204216}}]. [Erratum: Phys.Rev.D 48, 5965 (1993)].

\bibitem{Arnold:1992rz}
P.~B. Arnold and O.~Espinosa, {\it {The Effective potential and first order phase transitions: Beyond leading-order}},  {\em Phys. Rev. D} {\bf 47} (1993) 3546, [\href{http://arxiv.org/abs/hep-ph/9212235}{{\tt hep-ph/9212235}}]. [Erratum: Phys.Rev.D 50, 6662 (1994)].

\bibitem{Linde:1980tt}
A.~D. Linde, {\it {Fate of the False Vacuum at Finite Temperature: Theory and Applications}},  {\em Phys. Lett. B} {\bf 100} (1981) 37--40.

\bibitem{Adams:1993zs}
F.~C. Adams, {\it {General solutions for tunneling of scalar fields with quartic potentials}},  {\em Phys. Rev. D} {\bf 48} (1993) 2800--2805, [\href{http://arxiv.org/abs/hep-ph/9302321}{{\tt hep-ph/9302321}}].

\bibitem{Turner:1990rc}
M.~S. Turner and F.~Wilczek, {\it {Relic gravitational waves and extended inflation}},  {\em Phys. Rev. Lett.} {\bf 65} (1990) 3080--3083.

\bibitem{Kosowsky:1991ua}
A.~Kosowsky, M.~S. Turner, and R.~Watkins, {\it {Gravitational radiation from colliding vacuum bubbles}},  {\em Phys. Rev. D} {\bf 45} (1992) 4514--4535.

\bibitem{Kosowsky:1992rz}
A.~Kosowsky, M.~S. Turner, and R.~Watkins, {\it {Gravitational waves from first order cosmological phase transitions}},  {\em Phys. Rev. Lett.} {\bf 69} (1992) 2026--2029.

\bibitem{Kosowsky:1992vn}
A.~Kosowsky and M.~S. Turner, {\it {Gravitational radiation from colliding vacuum bubbles: envelope approximation to many bubble collisions}},  {\em Phys. Rev. D} {\bf 47} (1993) 4372--4391, [\href{http://arxiv.org/abs/astro-ph/9211004}{{\tt astro-ph/9211004}}].

\bibitem{Turner:1992tz}
M.~S. Turner, E.~J. Weinberg, and L.~M. Widrow, {\it {Bubble nucleation in first order inflation and other cosmological phase transitions}},  {\em Phys. Rev. D} {\bf 46} (1992) 2384--2403.

\bibitem{Hindmarsh:2013xza}
M.~Hindmarsh, S.~J. Huber, K.~Rummukainen, and D.~J. Weir, {\it {Gravitational waves from the sound of a first order phase transition}},  {\em Phys. Rev. Lett.} {\bf 112} (2014) 041301, [\href{http://arxiv.org/abs/1304.2433}{{\tt arXiv:1304.2433}}].

\bibitem{Giblin:2014qia}
J.~T. Giblin and J.~B. Mertens, {\it {Gravitional radiation from first-order phase transitions in the presence of a fluid}},  {\em Phys. Rev.} {\bf D90} (2014), no.~2 023532, [\href{http://arxiv.org/abs/1405.4005}{{\tt arXiv:1405.4005}}].

\bibitem{Hindmarsh:2015qta}
M.~Hindmarsh, S.~J. Huber, K.~Rummukainen, and D.~J. Weir, {\it {Numerical simulations of acoustically generated gravitational waves at a first order phase transition}},  {\em Phys. Rev.} {\bf D92} (2015), no.~12 123009, [\href{http://arxiv.org/abs/1504.03291}{{\tt arXiv:1504.03291}}].

\bibitem{Hindmarsh:2017gnf}
M.~Hindmarsh, S.~J. Huber, K.~Rummukainen, and D.~J. Weir, {\it {Shape of the acoustic gravitational wave power spectrum from a first order phase transition}},  {\em Phys. Rev. D} {\bf 96} (2017), no.~10 103520, [\href{http://arxiv.org/abs/1704.05871}{{\tt arXiv:1704.05871}}]. [Erratum: Phys.Rev.D 101, 089902 (2020)].

\bibitem{Kamionkowski:1993fg}
M.~Kamionkowski, A.~Kosowsky, and M.~S. Turner, {\it {Gravitational radiation from first order phase transitions}},  {\em Phys. Rev. D} {\bf 49} (1994) 2837--2851, [\href{http://arxiv.org/abs/astro-ph/9310044}{{\tt astro-ph/9310044}}].

\bibitem{Kosowsky:2001xp}
A.~Kosowsky, A.~Mack, and T.~Kahniashvili, {\it {Gravitational radiation from cosmological turbulence}},  {\em Phys. Rev. D} {\bf 66} (2002) 024030, [\href{http://arxiv.org/abs/astro-ph/0111483}{{\tt astro-ph/0111483}}].

\bibitem{Caprini:2006jb}
C.~Caprini and R.~Durrer, {\it {Gravitational waves from stochastic relativistic sources: Primordial turbulence and magnetic fields}},  {\em Phys. Rev.} {\bf D74} (2006) 063521, [\href{http://arxiv.org/abs/astro-ph/0603476}{{\tt astro-ph/0603476}}].

\bibitem{Gogoberidze:2007an}
G.~Gogoberidze, T.~Kahniashvili, and A.~Kosowsky, {\it {The Spectrum of Gravitational Radiation from Primordial Turbulence}},  {\em Phys. Rev. D} {\bf 76} (2007) 083002, [\href{http://arxiv.org/abs/0705.1733}{{\tt arXiv:0705.1733}}].

\bibitem{Caprini:2009yp}
C.~Caprini, R.~Durrer, and G.~Servant, {\it {The stochastic gravitational wave background from turbulence and magnetic fields generated by a first-order phase transition}},  {\em JCAP} {\bf 0912} (2009) 024, [\href{http://arxiv.org/abs/0909.0622}{{\tt arXiv:0909.0622}}].

\bibitem{Niksa:2018ofa}
P.~Niksa, M.~Schlederer, and G.~Sigl, {\it {Gravitational Waves produced by Compressible MHD Turbulence from Cosmological Phase Transitions}},  {\em Class. Quant. Grav.} {\bf 35} (2018), no.~14 144001, [\href{http://arxiv.org/abs/1803.02271}{{\tt arXiv:1803.02271}}].

\bibitem{Caprini:2015zlo}
C.~Caprini {\em et~al.}, {\it {Science with the space-based interferometer eLISA. II: Gravitational waves from cosmological phase transitions}},  {\em JCAP} {\bf 1604} (2016), no.~04 001, [\href{http://arxiv.org/abs/1512.06239}{{\tt arXiv:1512.06239}}].

\bibitem{Steinhardt:1981ct}
P.~J. Steinhardt, {\it {Relativistic Detonation Waves and Bubble Growth in False Vacuum Decay}},  {\em Phys. Rev.} {\bf D25} (1982) 2074.

\bibitem{Espinosa:2010hh}
J.~R. Espinosa, T.~Konstandin, J.~M. No, and G.~Servant, {\it {Energy Budget of Cosmological First-order Phase Transitions}},  {\em JCAP} {\bf 06} (2010) 028, [\href{http://arxiv.org/abs/1004.4187}{{\tt arXiv:1004.4187}}].

\bibitem{Lewicki:2021pgr}
M.~Lewicki, M.~Merchand, and M.~Zych, {\it {Electroweak bubble wall expansion: gravitational waves and baryogenesis in Standard Model-like thermal plasma}},  {\em JHEP} {\bf 02} (2022) 017, [\href{http://arxiv.org/abs/2111.02393}{{\tt arXiv:2111.02393}}].

\bibitem{Belyaev:2012qa}
A.~Belyaev, N.~D. Christensen, and A.~Pukhov, {\it {CalcHEP 3.4 for collider physics within and beyond the Standard Model}},  {\em Comput. Phys. Commun.} {\bf 184} (2013) 1729--1769, [\href{http://arxiv.org/abs/1207.6082}{{\tt arXiv:1207.6082}}].

\bibitem{Alguero:2023zol}
G.~Alguero, G.~Belanger, F.~Boudjema, S.~Chakraborti, A.~Goudelis, S.~Kraml, A.~Mjallal, and A.~Pukhov, {\it {micrOMEGAs 6.0: N-component dark matter}},  {\em Comput. Phys. Commun.} {\bf 299} (2024) 109133, [\href{http://arxiv.org/abs/2312.14894}{{\tt arXiv:2312.14894}}].

\bibitem{Crowder:2005nr}
J.~Crowder and N.~J. Cornish, {\it {Beyond LISA: Exploring future gravitational wave missions}},  {\em Phys. Rev. D} {\bf 72} (2005) 083005, [\href{http://arxiv.org/abs/gr-qc/0506015}{{\tt gr-qc/0506015}}].

\bibitem{Corbin:2005ny}
V.~Corbin and N.~J. Cornish, {\it {Detecting the cosmic gravitational wave background with the big bang observer}},  {\em Class. Quant. Grav.} {\bf 23} (2006) 2435--2446, [\href{http://arxiv.org/abs/gr-qc/0512039}{{\tt gr-qc/0512039}}].

\bibitem{Harry:2006fi}
G.~M. Harry, P.~Fritschel, D.~A. Shaddock, W.~Folkner, and E.~S. Phinney, {\it {Laser interferometry for the big bang observer}},  {\em Class. Quant. Grav.} {\bf 23} (2006) 4887--4894. [Erratum: Class. Quant. Grav.23,7361(2006)].

\bibitem{Seto:2001qf}
N.~Seto, S.~Kawamura, and T.~Nakamura, {\it {Possibility of direct measurement of the acceleration of the universe using 0.1-Hz band laser interferometer gravitational wave antenna in space}},  {\em Phys. Rev. Lett.} {\bf 87} (2001) 221103, [\href{http://arxiv.org/abs/astro-ph/0108011}{{\tt astro-ph/0108011}}].

\bibitem{Kawamura:2006up}
S.~Kawamura {\em et~al.}, {\it {The Japanese space gravitational wave antenna DECIGO}},  {\em Class. Quant. Grav.} {\bf 23} (2006) S125--S132.

\bibitem{Yagi:2011wg}
K.~Yagi and N.~Seto, {\it {Detector configuration of DECIGO/BBO and identification of cosmological neutron-star binaries}},  {\em Phys. Rev. D} {\bf 83} (2011) 044011, [\href{http://arxiv.org/abs/1101.3940}{{\tt arXiv:1101.3940}}]. [Erratum: Phys.Rev.D 95, 109901 (2017)].

\bibitem{Punturo:2010zz}
M.~Punturo {\em et~al.}, {\it {The Einstein Telescope: A third-generation gravitational wave observatory}},  {\em Class. Quant. Grav.} {\bf 27} (2010) 194002.

\bibitem{Hild:2010id}
S.~Hild {\em et~al.}, {\it {Sensitivity Studies for Third-Generation Gravitational Wave Observatories}},  {\em Class. Quant. Grav.} {\bf 28} (2011) 094013, [\href{http://arxiv.org/abs/1012.0908}{{\tt arXiv:1012.0908}}].

\bibitem{Sathyaprakash:2012jk}
B.~Sathyaprakash {\em et~al.}, {\it {Scientific Objectives of Einstein Telescope}},  {\em Class. Quant. Grav.} {\bf 29} (2012) 124013, [\href{http://arxiv.org/abs/1206.0331}{{\tt arXiv:1206.0331}}]. [Erratum: Class.Quant.Grav. 30, 079501 (2013)].

\bibitem{ET:2019dnz}
{\bf ET} Collaboration, M.~Maggiore {\em et~al.}, {\it {Science Case for the Einstein Telescope}},  {\em JCAP} {\bf 03} (2020) 050, [\href{http://arxiv.org/abs/1912.02622}{{\tt arXiv:1912.02622}}].

\bibitem{2017arXiv170200786A}
{\bf LISA} Collaboration, P.~Amaro-Seoane~et al, {\it {Laser Interferometer Space Antenna}},  {\em arXiv e-prints} (Feb., 2017) arXiv:1702.00786, [\href{http://arxiv.org/abs/1702.00786}{{\tt arXiv:1702.00786}}].

\bibitem{Sesana:2019vho}
A.~Sesana {\em et~al.}, {\it {Unveiling the gravitational universe at $\mu$-Hz frequencies}},  {\em Exper. Astron.} {\bf 51} (2021), no.~3 1333--1383, [\href{http://arxiv.org/abs/1908.11391}{{\tt arXiv:1908.11391}}].

\bibitem{Garcia-Bellido:2021zgu}
J.~Garcia-Bellido, H.~Murayama, and G.~White, {\it {Exploring the Early Universe with Gaia and THEIA}},  \href{http://arxiv.org/abs/2104.04778}{{\tt arXiv:2104.04778}}.

\bibitem{Schmitz:2020syl}
K.~Schmitz, {\it {New Sensitivity Curves for Gravitational-Wave Experiments}},  \href{http://arxiv.org/abs/2002.04615}{{\tt arXiv:2002.04615}}.

\bibitem{Gustafsson:2012aj}
M.~Gustafsson, S.~Rydbeck, L.~Lopez-Honorez, and E.~Lundstrom, {\it {Status of the Inert Doublet Model and the Role of multileptons at the LHC}},  {\em Phys. Rev. D} {\bf 86} (2012) 075019, [\href{http://arxiv.org/abs/1206.6316}{{\tt arXiv:1206.6316}}].

\bibitem{Datta:2016nfz}
A.~Datta, N.~Ganguly, N.~Khan, and S.~Rakshit, {\it {Exploring collider signatures of the inert Higgs doublet model}},  {\em Phys. Rev. D} {\bf 95} (2017), no.~1 015017, [\href{http://arxiv.org/abs/1610.00648}{{\tt arXiv:1610.00648}}].

\bibitem{Poulose:2016lvz}
P.~Poulose, S.~Sahoo, and K.~Sridhar, {\it {Exploring the Inert Doublet Model through the dijet plus missing transverse energy channel at the LHC}},  {\em Phys. Lett. B} {\bf 765} (2017) 300--306, [\href{http://arxiv.org/abs/1604.03045}{{\tt arXiv:1604.03045}}].

\bibitem{Miao:2010rg}
X.~Miao, S.~Su, and B.~Thomas, {\it {Trilepton Signals in the Inert Doublet Model}},  {\em Phys. Rev. D} {\bf 82} (2010) 035009, [\href{http://arxiv.org/abs/1005.0090}{{\tt arXiv:1005.0090}}].

\bibitem{Belyaev:2016lok}
A.~Belyaev, G.~Cacciapaglia, I.~P. Ivanov, F.~Rojas-Abatte, and M.~Thomas, {\it {Anatomy of the Inert Two Higgs Doublet Model in the light of the LHC and non-LHC Dark Matter Searches}},  {\em Phys. Rev. D} {\bf 97} (2018), no.~3 035011, [\href{http://arxiv.org/abs/1612.00511}{{\tt arXiv:1612.00511}}].

\bibitem{Belyaev:2018ext}
A.~Belyaev, T.~R. Fernandez Perez~Tomei, P.~G. Mercadante, C.~S. Moon, S.~Moretti, S.~F. Novaes, L.~Panizzi, F.~Rojas, and M.~Thomas, {\it {Advancing LHC probes of dark matter from the inert two-Higgs-doublet model with the monojet signal}},  {\em Phys. Rev. D} {\bf 99} (2019), no.~1 015011, [\href{http://arxiv.org/abs/1809.00933}{{\tt arXiv:1809.00933}}].

\bibitem{KATRIN:2019yun}
{\bf KATRIN} Collaboration, M.~Aker {\em et~al.}, {\it {Improved Upper Limit on the Neutrino Mass from a Direct Kinematic Method by KATRIN}},  {\em Phys. Rev. Lett.} {\bf 123} (2019), no.~22 221802, [\href{http://arxiv.org/abs/1909.06048}{{\tt arXiv:1909.06048}}].

\bibitem{Liu:2025xvm}
W.~Liu and Y.~Wu, {\it {Testing Leptogenesis from Observable Gravitational Waves}},  \href{http://arxiv.org/abs/2504.07819}{{\tt arXiv:2504.07819}}.

\bibitem{Caprini:2019egz}
C.~Caprini {\em et~al.}, {\it {Detecting gravitational waves from cosmological phase transitions with LISA: an update}},  {\em JCAP} {\bf 03} (2020) 024, [\href{http://arxiv.org/abs/1910.13125}{{\tt arXiv:1910.13125}}].

\bibitem{Guo:2020grp}
H.-K. Guo, K.~Sinha, D.~Vagie, and G.~White, {\it {Phase Transitions in an Expanding Universe: Stochastic Gravitational Waves in Standard and Non-Standard Histories}},  \href{http://arxiv.org/abs/2007.08537}{{\tt arXiv:2007.08537}}.

\end{thebibliography}
\providecommand{\href}[2]{#2}\begingroup\raggedright\endgroup

\end{document}